\documentclass{PoS}
\usepackage{amsmath}
\usepackage{amsfonts}
\usepackage{graphicx}
\usepackage{bm}
\usepackage[ragged]{sidecap}

\def \be {\begin{equation}}
\def \ee {\end{equation}}

\title{Regge Amplitudes for Two-to-Two Reactions\thanks{Worked supported by Indiana University Collaborative Research Grant (IUCRG)}}

\ShortTitle{Regge Amplitudes for Two-to-Two Reactions}

\author{\speaker{Vincent Mathieu}\\
        	ECT*, Villa Tambosi, I-38123 Villazzano (Trento), Italy\\
	Physics Department, Indiana University, Bloomington, IN 47405, USA\\
	Center For Exploration of Energy and Matter, Indiana University, Bloomington, IN 47408, USA\\
        	E-mail: \email{vincent.mathieu@umons.ac.be}}

\abstract{We present a fit based on Regge theory of two-to-two reactions at high energies particulary focused on leading non-strange positive naturality exchanges. Factorization of the residues is assumed between beam and target vertices. This study is a first step toward the analysis of multiple mesons peripherical production.}

\FullConference{QCD-TNT-III-From quarks and gluons to hadronic matter: A bridge too far?,\\
		2-6 September, 2013\\
		European Centre for Theoretical Studies in Nuclear Physics and Related Areas (ECT*), Villazzano, Trento (Italy)}

\begin{document}

\section{Introduction}

A quantitative description of the hadron spectrum is essential for a complete understanding of Quantum Chromodynamics (QCD), the theory of the strong interactions. Numerous experiments devoted to hadron spectroscopy are currently underway {\it e.g.} COMPASS and BESIII, or are planned for the near future {\it e.g.} GlueX and CLAS12 at Jefferson Lab, and PANDA at GSI. Those new generations of high statistics and precision experiments demand a level of detailed partial wave decomposition and amplitude analysis never achieved before. 



Extracting properties of weakly coupled resonances requires a detailed understanding of the background. For instance, the so-called Deck mechanism based on Regge theory interferes with the production of the $\pi_1$ in the $3\pi$ channel \cite{Nerling:2012ei}. Complex angular momentum theory, Regge theory, alone does not predict the parameters of such mechanism. In order to parametrize multiple mesons production amplitudes, one need first to determine the parameters on elementary reactions. We present here a summary of a detailed study on 2-to-2 processes. The details of the parametrizations, the fitting procedure and the parameters will be published elsewhere \cite{paper}. Since we want to export the knowledge gained to more complicated reactions, we will assume factorisation of the residues, {\it i.e.}  the global magnitude of the amplitude is split into a coupling at the target vertex and another coupling at the beam vertex. 

In Section \ref{sec:regge}, we discuss the specifications of the Regge pole and cuts. The interested readers will find more details and original references in the textbooks \cite{Collins:1971ff,Drechsler:1970mq}. We then start the analysis with pseudoscalar-nucleon and nucleon-nucleon processes in Section \ref{sec:pionnucl} and Section  \ref{sec:nuclnucl}. Photoproduction of pseudoscalars and vector production are detailed in Section \ref{sec:phot}. The conclusions are presented in Section \ref{sec:concl}. 
 
\section{Regge Poles and Cuts} \label{sec:regge}

Let us consider a given reaction $1+2\rightarrow 3+4$, called $s-$channel process. The corresponding $t-$ and $u-$channels are defined by $1+\bar 3\rightarrow \bar 2+4$ and $1+\bar 4\rightarrow \bar 2+ 3$. The crossing hypothesis assumes the three channels are related by the same analytical function $A(s,t,u)$ of the complex $(s,t,u)$ variables evaluated in three different domains. Singularities of analytical functions are poles and branch cuts. In a physical region, poles correspond to bound-states and resonances and multiple scatterings or cascade of decaying resonances produce cuts. Moreover the unitarity of the $S-$matrix restricts the singularities to physical regions. The knowledge of all singularities in the whole domain of an analytical function could in principle determines his value in any point of the domain. This idea led to dispersive relations widely used in hadronic physics. Another idea originally proposed by Regge is to consider partial wave amplitudes as analytical functions of the complex angular momentum.  Consequently, as explained in \cite{Collins:1971ff,paper}, the amplitude is a given channel is controlled by the singularities in the two other channels. 

For a Regge pole of trajectory $\alpha$ and signature $\tau$ we assume the form
\be\label{pole}
R(\nu,\alpha,\tau) = \beta(t) \kappa(\tau,\alpha) \Gamma(j_0-\alpha)(\nu/\nu_0)^{\alpha},
\ee
Note that $R(\alpha,\tau)$ is dimensionless by construction, $\nu=(s-u)/2$ is the crossing variable and $\nu_0=1$ GeV$^2$. In the parametrization \eqref{pole} appears the $j_0$ the lowest spin of the physical particles on the trajectory, {\it i.e.} $j_0=1$ for $(\rho,\omega,b,h)$ and $j_0=2$ for $(a,f)$. For the Pomeron a better description of the data is achieved for $j_0=1$ although no physical particle is know on this trajectory. It is often argued that glueballs would lie on the Pomeron trajectory \cite{Meyer:2004gx,Meyer:2004jc,Mathieu:2008me} but no definitive conclusion has been drawn yet. The signature factor is
$
\kappa(\tau,\alpha) = \frac{1}{2}(1+\tau e^{-i\pi\alpha}).
$

Regge trajectories are labelled by their signature $\tau=(-1)^J$ and naturality $\eta=P(-1)^J$. For the specific case of strong interactions, QCD is the underlying theory and provides additional symmetries. We will consider isospin as an exact symmetry and classify the trajectories also  according their isospin $I$ and $G-$parity $G=C(-1)^I$. 
We focus our attention on non-strange trajectories $I=0,1$ and out of the 16 possibilities for $(I,G,\tau,\eta)$ we keep only the height leading trajectories $(\omega,\rho,f,a)$ for natural exchanges and $(\pi,\eta,b,h)$ for unnatural exchanges, see Table \ref{tab:regge}. For a given set $(I,G,\tau,\eta)$ can correspond many trajectories, the leading one and its daughters trajectories.

\begin{table}[h]\caption{Non-strange Regge Trajectories and Regge exchanges for elastic and inelastic scatterings. \label{tab:regge}}
\begin{tabular}{| c | c  || c | c | | l| l|| l| l|}
\hline
$I^{G\tau\eta}$ & &$I^{G\tau\eta}$ & &&&&\\
\hline
$0^{+++}$ &$f$ &   $0^{+--}$ &$\bar f$ & $\pi^\pm p$ & $\mathbb{P} + f \pm \rho$ &$\pi^-p\to\pi^0 n$ & $\sqrt{2}\rho$\\
$0^{--+}$ &$\omega$ &  $0^{-+-}$ &$\bar\omega$ & $\pi^\pm n$ & $\mathbb{P} + f \mp \rho$ &$\pi^-p\to \eta^{(')} n$ & $\sqrt{2}a$ \\
$1^{-++}$ &$a$ & $1^{---}$ &$\bar a$ & $K^\pm p$ & $\mathbb{P} + f \pm \rho + a \pm \omega$ & $K^+ n\to K^0 p$ & $ \sqrt{2}(\rho + a)$\\
$1^{+-+}$ &$\rho$ & $1^{++-}$ &$\bar\rho$ & $K^\pm n$ & $\mathbb{P} + f \mp \rho - a \pm \omega$ & $K^- p\to \bar K^0 n$ & $ \sqrt{2}(\rho + a)$\\
\hline
$0^{++-}$ &$\eta$ &$0^{+-+}$ &$\bar\eta$ & $pp$ & $\mathbb{P} + f + \rho + a + \omega $ (+ unnat) & $\gamma p\to\pi^0 p$ & $\omega+\rho+b+h$ \\
$0^{---}$ &$h$ &$0^{-++}$ &$\bar h$ & $\bar pp$ & $\mathbb{P} + f - \rho + a - \omega $  (+ unnat) & $\gamma n\to\pi^0 n$ & $\omega-\rho-b+h$\\
$1^{-+-}$ &$\pi$ &$1^{--+}$ &$\bar \pi$ & $pn$ & $\mathbb{P} + f - \rho - a + \omega $ (+ unnat) & $\pi^- p\to \omega n $ & $\rho+b$\\
$1^{+--}$ &$b$ &$1^{+++}$ &$\bar b$ & $\bar pn$ & $\mathbb{P} + f + \rho - a - \omega $  (+ unnat) && 
\\
\hline
\end{tabular}
\end{table}

Know resonances lie on linear trajectories. In the constituent quark model, two quarks linked by a color flux tube of tension $\sigma$ give indeed a linear relation between spin and mass squared with a universal slope $\alpha'=(2\pi \sigma)^{-1}\sim 0.9$ GeV$^{-2}$. We then parametrized the height trajectories by the simple form $\alpha(t) = \alpha_0+\alpha't$ with $\alpha'=0.9$ for $(\omega,\rho,f,a)$. However for unnatural mesons $(\pi,\eta,b,h)$ a slope $\alpha'\sim 0.7$ seem more appropriate.

Experimentally all total cross sections rise. This fact can be explained by the introduction of the Pomeron trajectory having vacuum quantum numbers as for the $f$ and an intercept greater than one. We fixed its intercept to the common value $\alpha_{\cal P}(0) = 1.08$. 

In the expression \eqref{pole}, the residues $\beta(t)$ will be fitted on data. We will test the hypothesis of factorability of the residues. The residue is split into a product of two couplings $\beta(t) = \beta_{ac}(t) \beta_{bd}(t)$. This hypothesis is essential to transpose the formalism to other reactions and make predictions for many meson production processes. The $t-$dependence of the couplings is weak but in some cases we will need a form $\beta(t)=\beta(0)e^{bt}$ inspired by an absorption model. The parameters will be published with the detailed analysis in Ref \cite{paper}.

Beside the poles, branching cuts also contribute in the dispersion relation. They correspond to the exchange and two distinct poles and can be parametrized by the same form \eqref{pole} but the overall magnitude is divided by $\log (\nu/\nu_0)$ according to the absorption model. The biggest contribution is provided by the exchange of the pole with the higher intercept, the Pomeron. We approximate the cut trajectory associated to a pole $\alpha(t) = \alpha_0+\alpha't$ to $\alpha_c(t) = \alpha_{0c}+\alpha^\prime_c t$ with $\alpha_{0c}=\alpha_0$ and $\alpha^\prime=\alpha^\prime({\mathbb P}) \alpha'/(\alpha^\prime({\mathbb P})+ \alpha')$ for simplicity. There is no factorization of the residue for the cuts. In principle cuts can be computed only from the knowledge of the pole. Here however we simply fit its residue on data (when a cut is needed) we the form
\be\label{cut}
R_c(\nu,\alpha,\tau) = \beta(t)[\log (\nu/\nu_0)]^{-1} \kappa(\tau,\alpha) \Gamma(j_0-\alpha)(\nu/\nu_0)^{\alpha}.
\ee

\section{Pseudoscalar-Nucleon Scatterings} \label{sec:pionnucl}
For equal masses scattering let $(M,\mu)$ be the masses of the nucleon and the pseudoscalar.
The amplitude is generally expressed in term of the two invariant amplitudes $(A,B)$. The helicity amplitudes are computed with ($H_i=s,t,u$ helicities)
\begin{align}
T_{H_i} & = \bar u_{H_i} \left[ A + {\scriptstyle\frac{1}{2}} (p_1+p_3)^\mu  \gamma_\mu B \right]u_{H_i}.
\end{align}
\begin{SCfigure}
	 \includegraphics[width=0.35\linewidth]{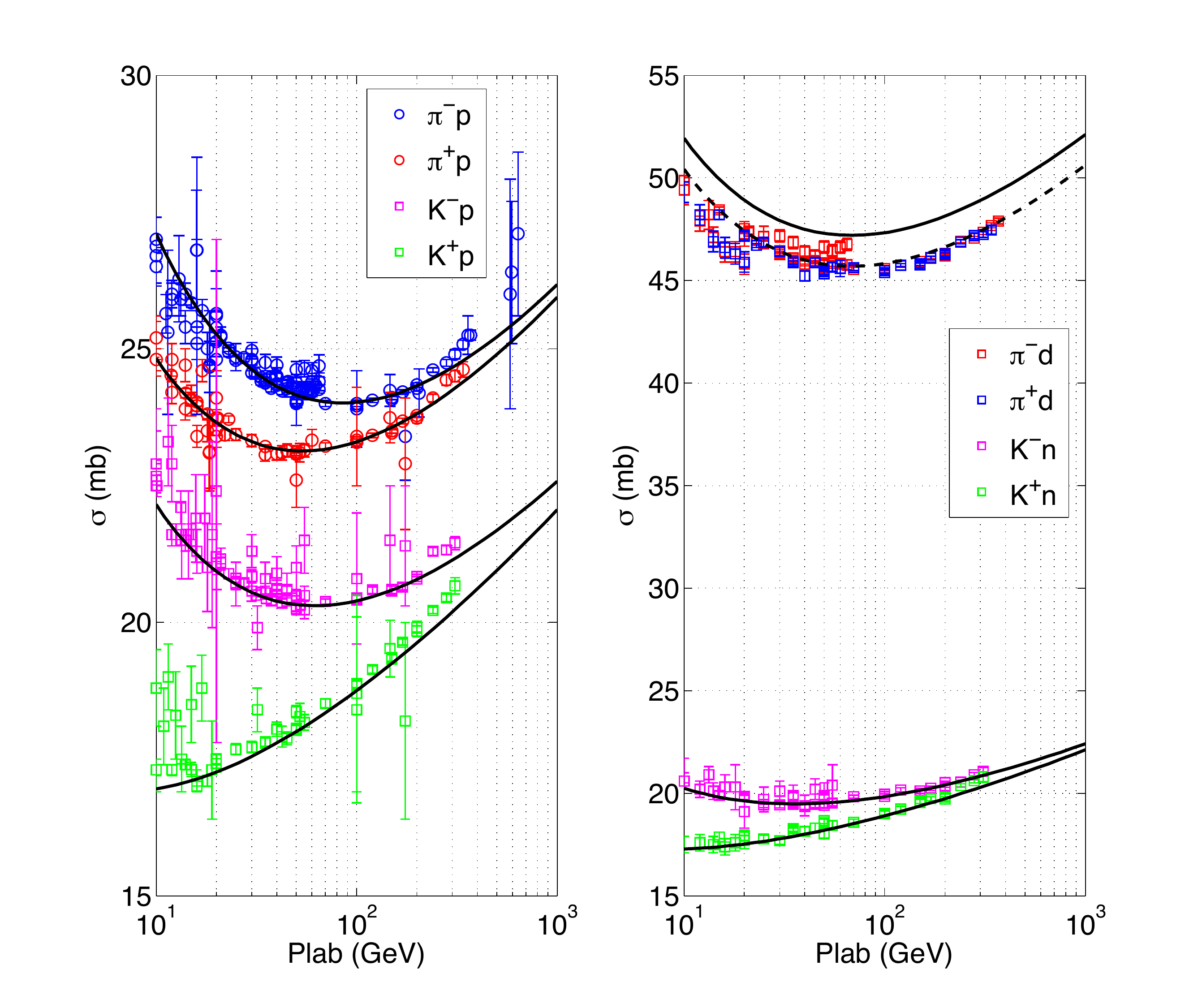} 
       \includegraphics[width=0.35\linewidth]{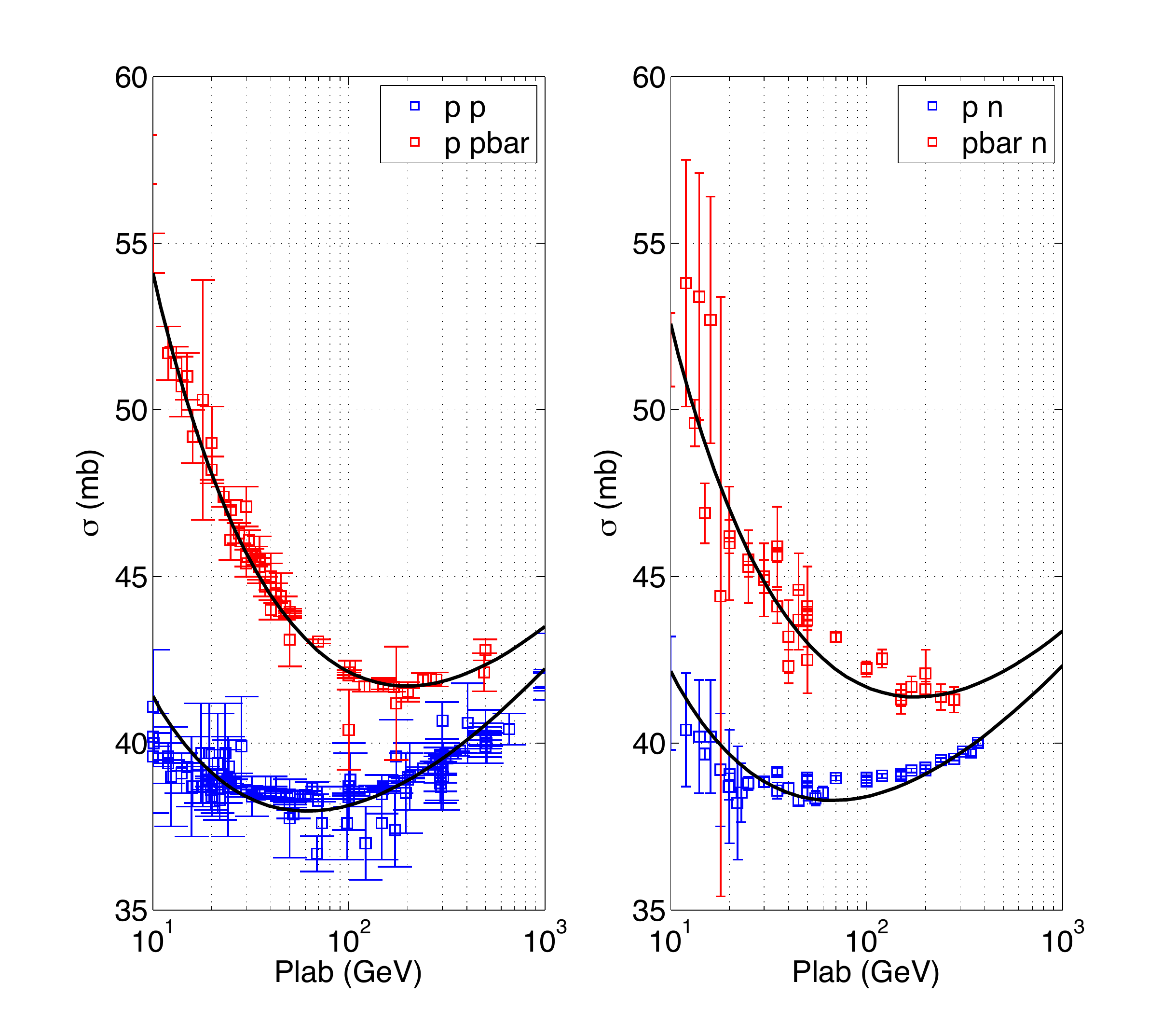} 
\caption{Total cross sections from left to right: a) $\pi^- p$ (blue) ; $\pi^+p$ (red) ; $K^- p$ (magenta) ; $K^+ p$ (green). b) $K^- n$ (magenta) ; $K^+ n$ (green).  c) $\bar p p$ (red) ; $pp$ (blue). d) $\bar p n$ (red) ; $pn$ (blue). Data from Particle Data Group \cite{Beringer:1900zz}. \label{fig:sigtot}}
\end{SCfigure}
We are interested in the total cross section $\sigma_{\text{tot}} = \sigma(12\to X)$, the differential cross section and the analyzing power. The observables read (remember the flux factor is $F_I = m_2 p_{\text{lab}}$)
\begin{subequations}
\begin{align}
\sigma_{\text{tot}}(s)&=  \frac{389.3}{2 F_I}\ [2MA+\nu B](s,0) && \text{in $\mu$b}\\
\frac{d\sigma}{dt} (s,t) &=  \frac{389.3\  }{64 \pi F_I^2}\left(|2MA+\nu B|^2 -t |A|^2\right) && \text{in $\mu$b.GeV}^{-2}  \\
2\text{Im}\,\rho^s_{+-}(s,t) &= \frac{2\sqrt{-t} \mathop{\text{Im}}\left[(2MA+\nu B)  A^*\right]}{ |2MA+\nu B|^2 -t |A|^2} .
\end{align}
\end{subequations}

There are two couplings at the nucleon vertex, helicity flip and non-flip. It is natural to associated the helicity non flip to the coupling in the forward direction. Therefore we define 
\begin{align}
F_n^+ &= 2MA+\nu B = \sum_e \beta^e_{00}(t)\beta^e_{++}(t) \kappa(\tau_e,\alpha_e) \Gamma(j_0-\alpha_e) (\nu/\nu_0)^{\alpha_e},\\
F^+_f &=  2M A = \sum_e \beta^e_{00}(t)\beta^e_{+-}(t) \kappa(\tau_e,\alpha_e) \Gamma(j_0-\alpha_e) (\nu/\nu_0)^{\alpha_e}. 
\end{align}
The two dimensionless functions $F^+_{n,f}$  involve a sum over all Regge poles and cuts [remember that for a cut there is an additional factor $\log^{-1}(\nu/\nu_0)$].

We already determined the Regge trajectories by matching the physical mesons on linear trajectories for $t>0$. We need now to fit the couplings. The non-flip couplings are fitted on total cross sections. We considered the 10 elastic reactions $\pi^\pm p, K^\pm p$, $K^\pm n$, $pp$, $\bar p p$, $pn$ and $\bar p n$. We do not include unnatural parity exchanges for the nucleon-nucleon scattering since they are sub-sub-leading.  The fit obtained for the total cross sections is consistent with \cite{Donnachie:1992ny}. The results of the fits for total cross sections are presented in Fig. \ref{fig:sigtot}. We give also the total cross section of pion scattering on deuteron target. The $\pi^\pm n$ is parameter free and the theoretical curve is exactly 1.5 mb above the experimental one. We shifted (the dashed line) the theoretical curve by this amount representing the binding energy of the two nuclei. 

\begin{figure}[h]
\begin{center}
	 \includegraphics[width=0.49\linewidth]{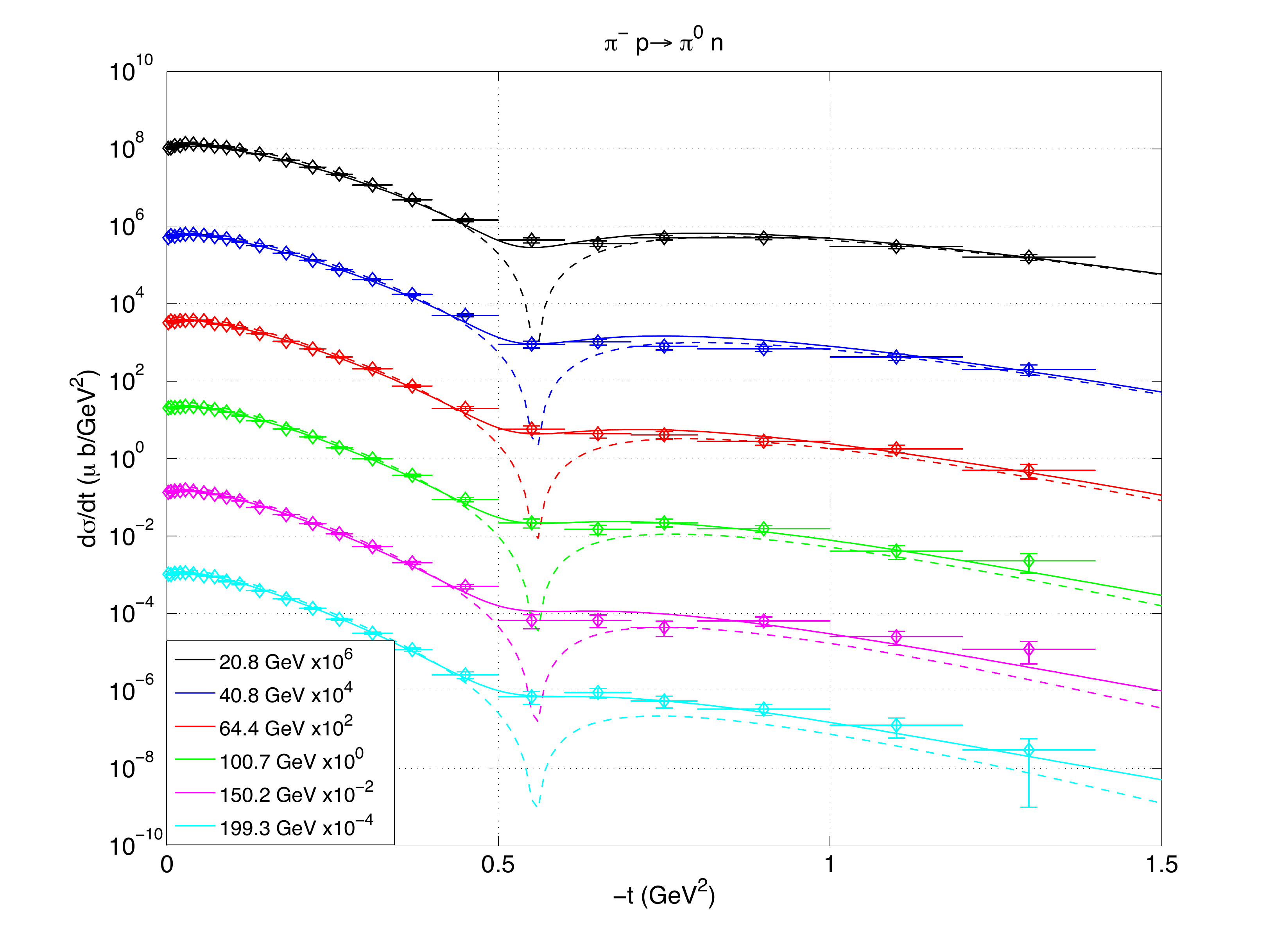} 
	 \includegraphics[width=0.45\linewidth]{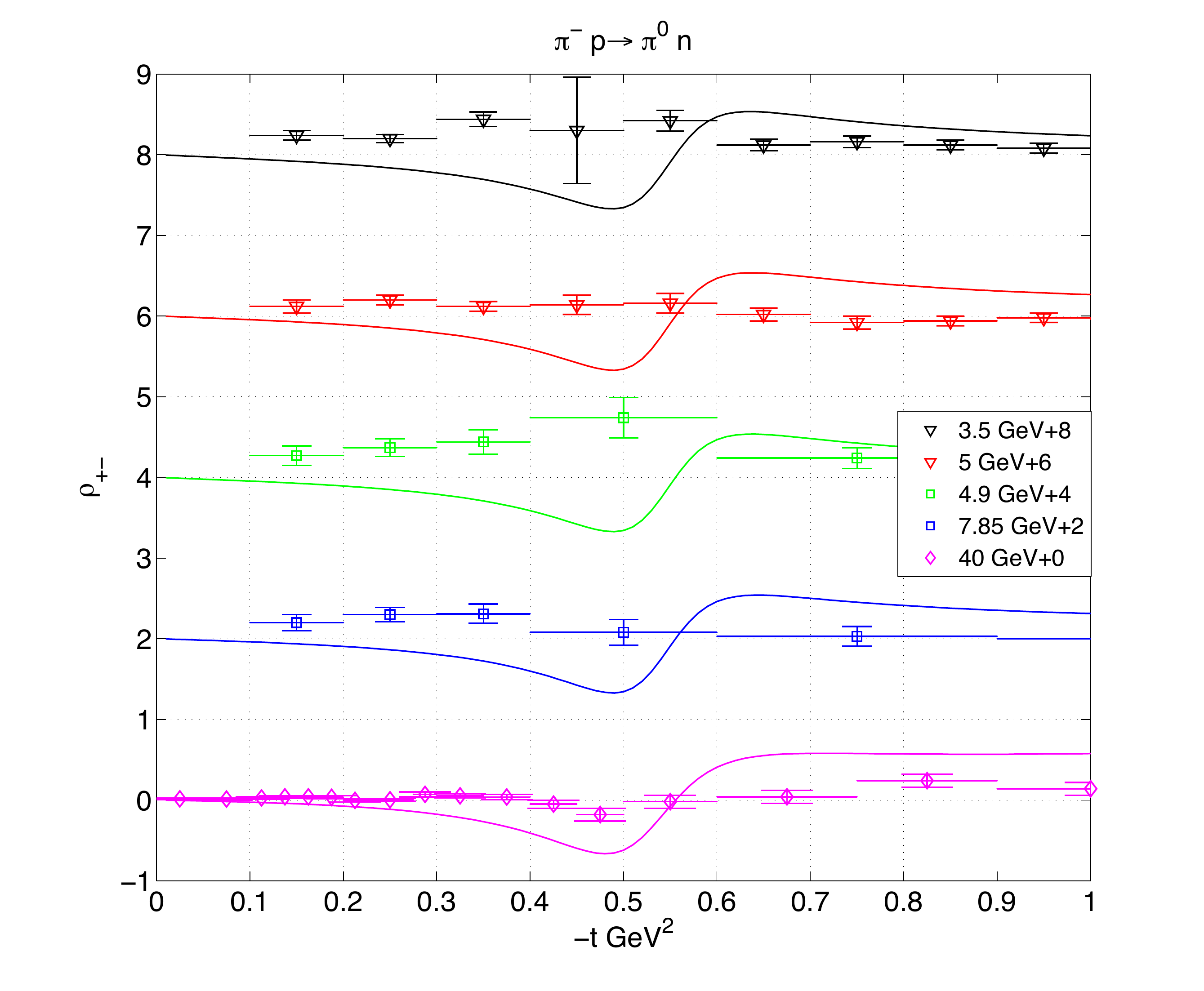} 
\end{center}
\caption{Left: $\pi^- p\to \pi^0 n$ differential cross section from $p_{\text{lab}}$=20.8 GeV (black) to  $p_{\text{lab}}$=199.3 GeV (cyan). Scaling factors are indicated on the figure. Theoretical models including pole and cut in solid lines and with only the $\rho$ pole in dashed lines. Data from \cite{Barnes:1976ek}. Right: Polarization $\rho_{+-}$ for $\pi^- p\to\pi^0 n$. Data from \cite{Bonamy:1973dz,Apokin:1986zc,Hill:1973bq}.\label{fig:Pi0P}}
\end{figure}

The helicity flip couplings are determined from differential cross sections. We first investigate processes involving only one and two exchanges. These are the charge exchanges reaction given in Table \ref{tab:regge}. The $\rho$ and $a$ trajectory have a wrong signature zero at respectively $t\sim-0.55$ GeV$^2$ and $t\sim -1.4$ GeV$^2$. They correspond to the zeros of the signature factor $1\pm e^{-i\pi\alpha}$. A dip is clearly seen in the data for $\pi^0$ production in Fig. \ref{fig:Pi0P}. Together with the energy dependence in the forward direction controlled by the intercept, we have a confirmation of the two parameters of the trajectory. However the differential cross section is not exactly zero and a better description of the data can be achieved with a cut. Its couplings is fitted on the cross section to fill the gap at $t\sim-0.55$ GeV$^2$.

The energy behavior of the differential cross section in the forward direction ($t=0$) allows us to determine the intercept of the $a$ trajectory and its coupling to $\pi\eta$ and $\pi\eta^\prime$. By comparing to the other charge exchange process $\pi^- p\to \pi^0 n$ involving only the $\rho$ pole we find different intercepts $\alpha_a(0)=0.4$ and $\alpha_\rho(0)=0.5$.  The ratio between $\pi\eta$ and $\pi\eta^\prime$  couplings could be interpreted as the tangent of the mixing angle. We find $\tan\phi=0.75$, {\it i.e.} $\phi\sim37^\circ$ not too far from the common value \cite{Feldmann:1998vh,Mathieu:2010ss}. The helicity flip coupling of the $a$ pole to the nucleon is fitted on the cross section.

\begin{SCfigure}
	 \includegraphics[width=0.35\linewidth]{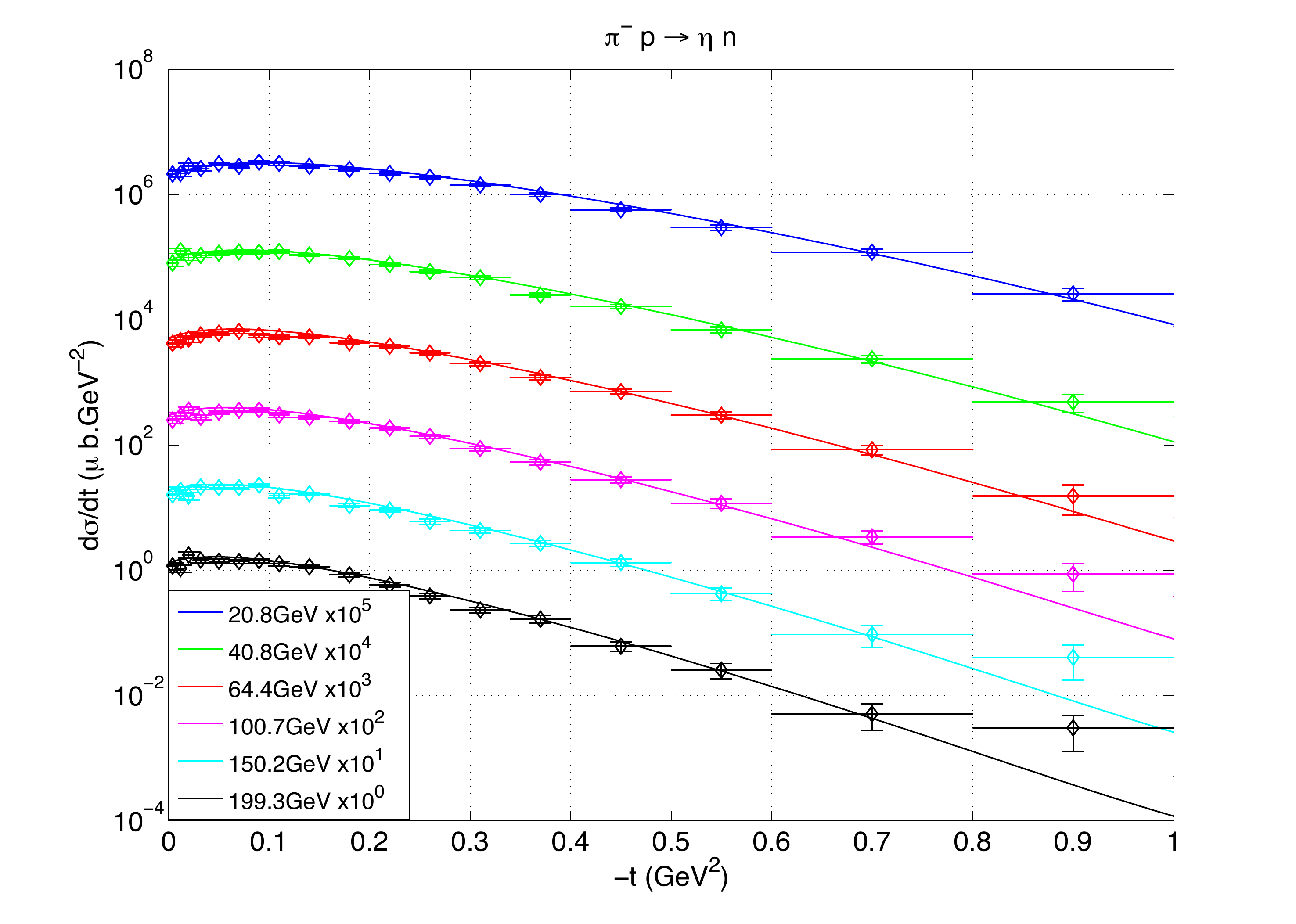}  
	 \includegraphics[width=0.35\linewidth]{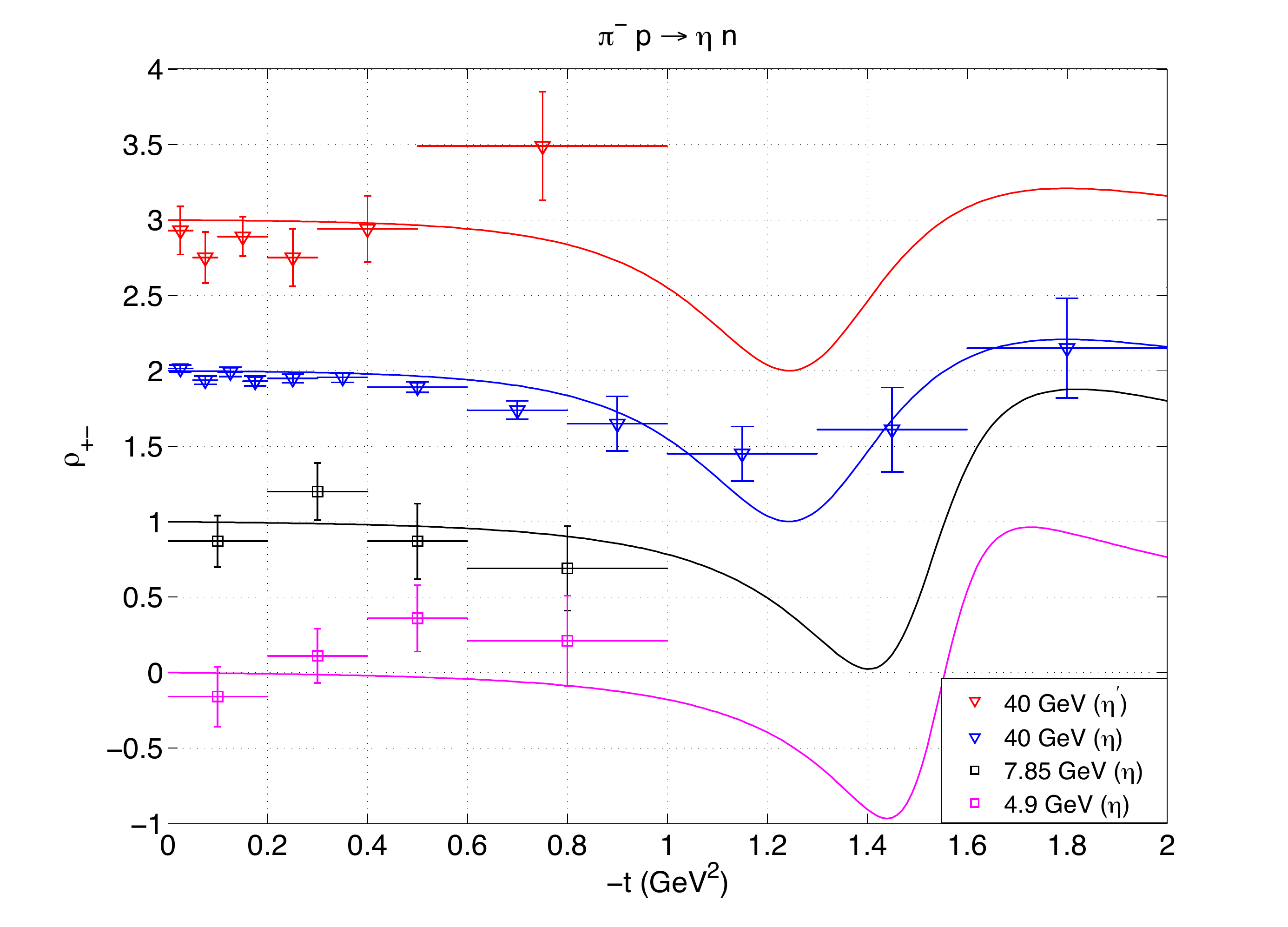} 
\caption{{\it Left}: $\pi^- p\to \eta n$ differential cross section.  Scaling factors are indicated. Data from \cite{Dahl:1976em} ; {\it right}: Polarization $\rho_{+-}$ for $\pi^- p\to \eta n$  and $\pi^- p\to \eta^\prime n$. Data from \cite{Apokin:1985cz,Bonamy:1973dz}.\label{fig:SigEta}}
\end{SCfigure}

The polarization observable is proportional to the imaginary part of $T_{++}T_{+-}^*$. The complex phase of an amplitude is coming from the signature factor $\kappa(\tau,\alpha)=(1/2)(1+\tau e^{-i\pi \alpha})$. The polarization then probes the interference between different contributions. A single pole does not contribute to this measurement. In the charge exchange reactions $\pi^-p\to(\pi^0,\eta,\eta^\prime)n$ only one signature is allowed for the leading trajectory. The polarization is sensitive to the interference between the pole and the cut (or a daughter trajectory but we did not investigate this case). Im($\rho_{+-}$) would changes sign for $t\sim-0.5$ GeV$^2$ in the presence of a cut in addition to the $\rho$ pole. However the polarization in Fig. \ref{fig:Pi0P} is always positive at low energies and changes sign for $t\sim-0.5$ GeV$^2$ only at high energies. This is probably causes by a secondary pole, the $\bar \rho$ pole with a positive naturality. Indeed the interference between $\rho$ and $\bar \rho$ is positive for $|t|<1.5$ GeV$^2$. But as the energy increases $\bar\rho$ pole, having the smaller intercept, disappear and remains only the interference between the $\rho$ pole and its cut. Note that in the solid lines in the polarization for $\pi^- p\to \pi^0 n$ is the model where the parameters were fitted only on the cross section and include only the $\rho$ pole and its cut. 

For the reaction $\pi^-p\to\eta n$ Im($\rho_{+-}$) would change sign for $t\sim-1.5$ GeV$^2$ in the presence of a cut in addition to the $a$ pole. This is exactly what is measured at $40$ GeV in $\eta$ production in Fig. \ref{fig:SigEta}.  We added a small cut for illustration but a pole only is sufficient for a very good description of the differential cross section. Note that we cannot explain the change of sign for $t\sim-0.5$ GeV$^2$ in $\eta^\prime$ production at $40$ GeV maybe caused by daughter trajectories (the $a_1(1260)$ lies on the $\bar a$ trajectory and is not included here).

\begin{figure}[htb]
\begin{center}
	 \includegraphics[width=0.4\linewidth]{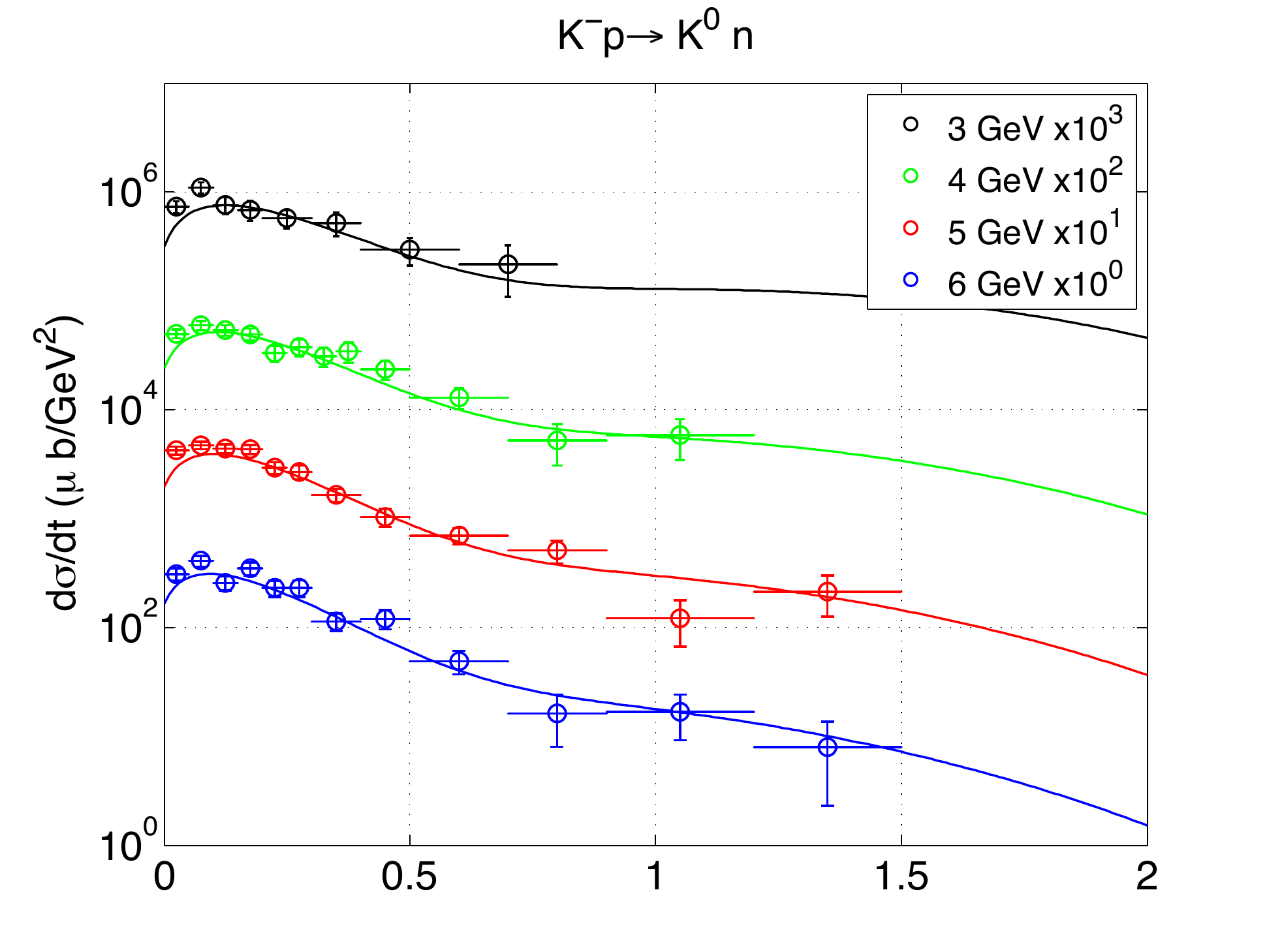}  
	 \includegraphics[width=0.4\linewidth]{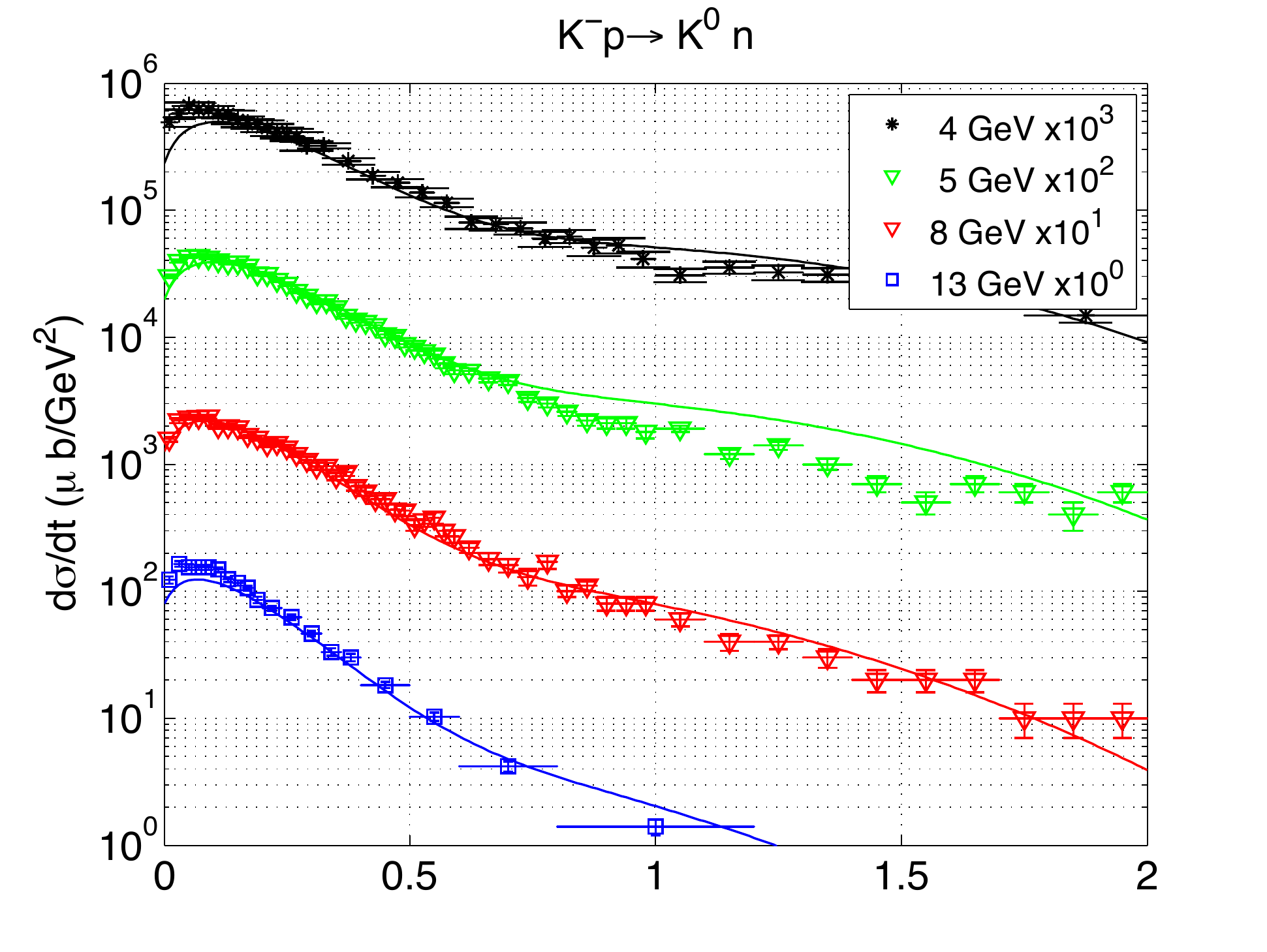} 
	 \includegraphics[width=0.4\linewidth]{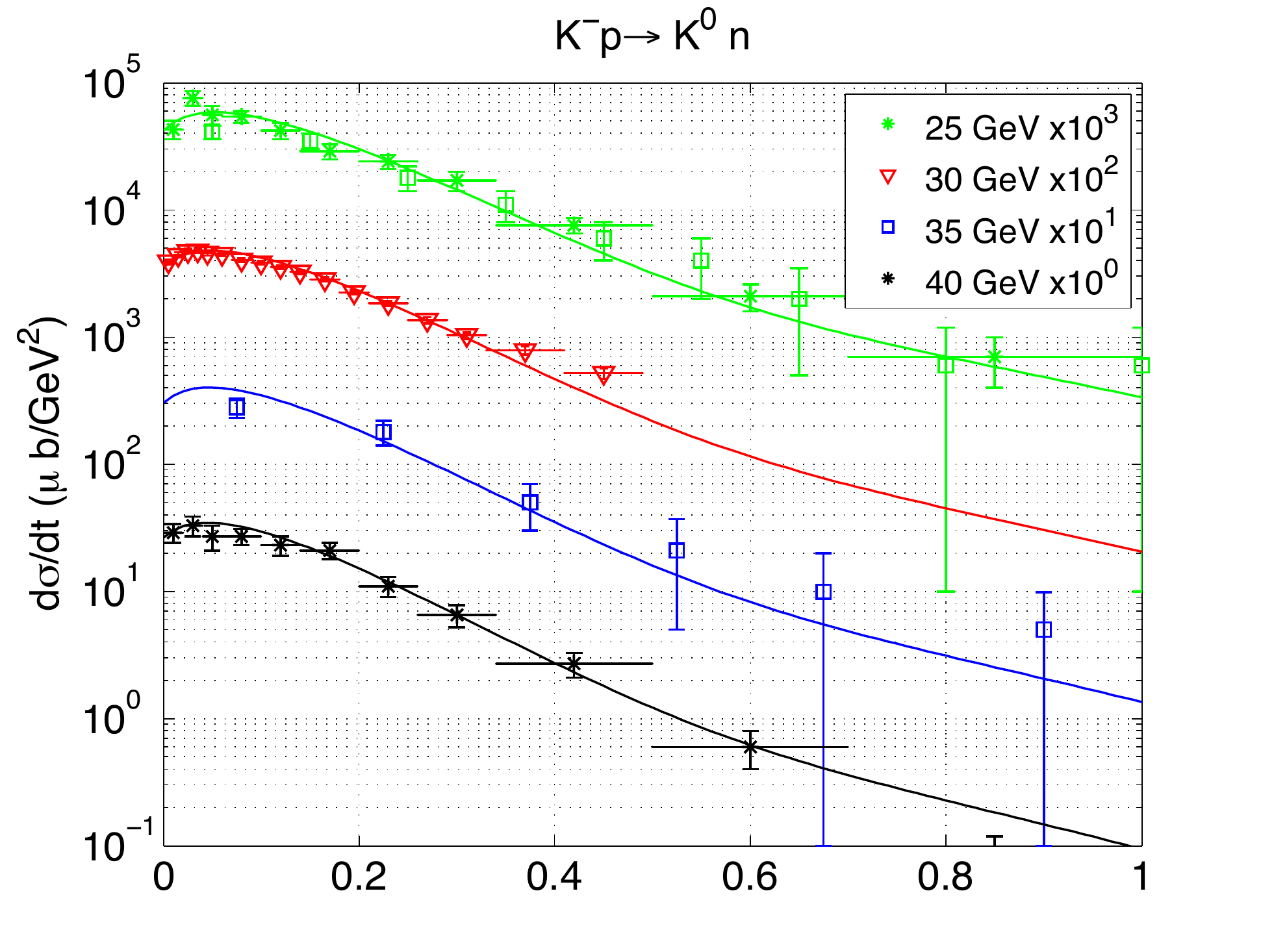}  
	 \includegraphics[width=0.4\linewidth]{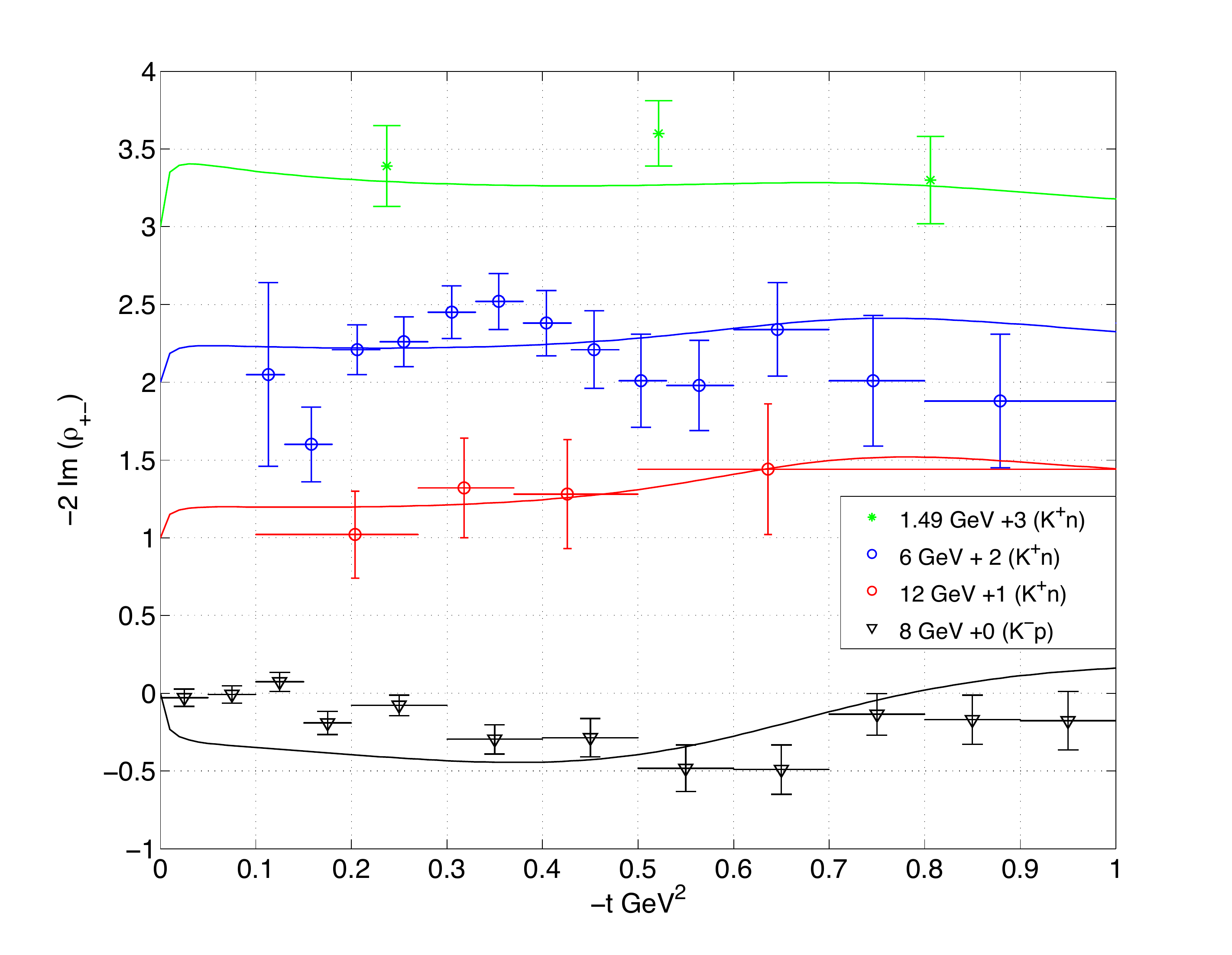} 
\end{center}
\caption{Top left, top right and bottom left: $K^- p\to \bar K^0 n$ differential cross section. Data from \cite{Ambats:1973hv,Marzano:1977fa,Gallivan:1976cf,Brandenburg:1976yj,Apel:1977ki, Binon:1980ud,Bolotov:1974pr}. 
Bottom right: Polarization -2Im($\rho_{+-}$) for $K^+ n\to K^0 p$ (green, blue and red) and $K^- p\to \bar K^0 n$ (black). Data from \cite{Nakajima:1982tm,Fujisaki:1979fm,Beusch:1974xk}.\label{fig:SigKN}}
\end{figure}

Having determined the helicity flip couplings to the nucleon of the $\rho$ and $a$ trajectories, we can compare the model predictions to the data for $K^+ n\to K^0 p$ and $K^- p\to \bar K^0 n$. All parameters are known since helicity non flip to the nucleon and $K\bar K$ couplings were obtained from the total elastic cross sections. The differential cross section agrees well with the data for a large spectrum of incident momentum energies as can be seen on Fig. \ref{fig:SigKN} for $p_{\text{lab}}\in[3,40]$ GeV. Although the polarization Im($\rho_{+-}$) do not present all the oscillations observed in the data on Fig \ref{fig:SigKN} they have the correct sign. The analyzing power is much more sensitive to a small correction to the pole approximation of the amplitude than the cross section as we saw in the charge exchange reactions $\pi^- p\to (\pi^0,\eta) n$.  The complicated structures in Im($\rho_{+-}$) could be produced by a small cut or a small interference with daughter trajectories (again the $\bar a$ pole was ignored).

\begin{SCfigure}
	 \includegraphics[width=0.3\linewidth]{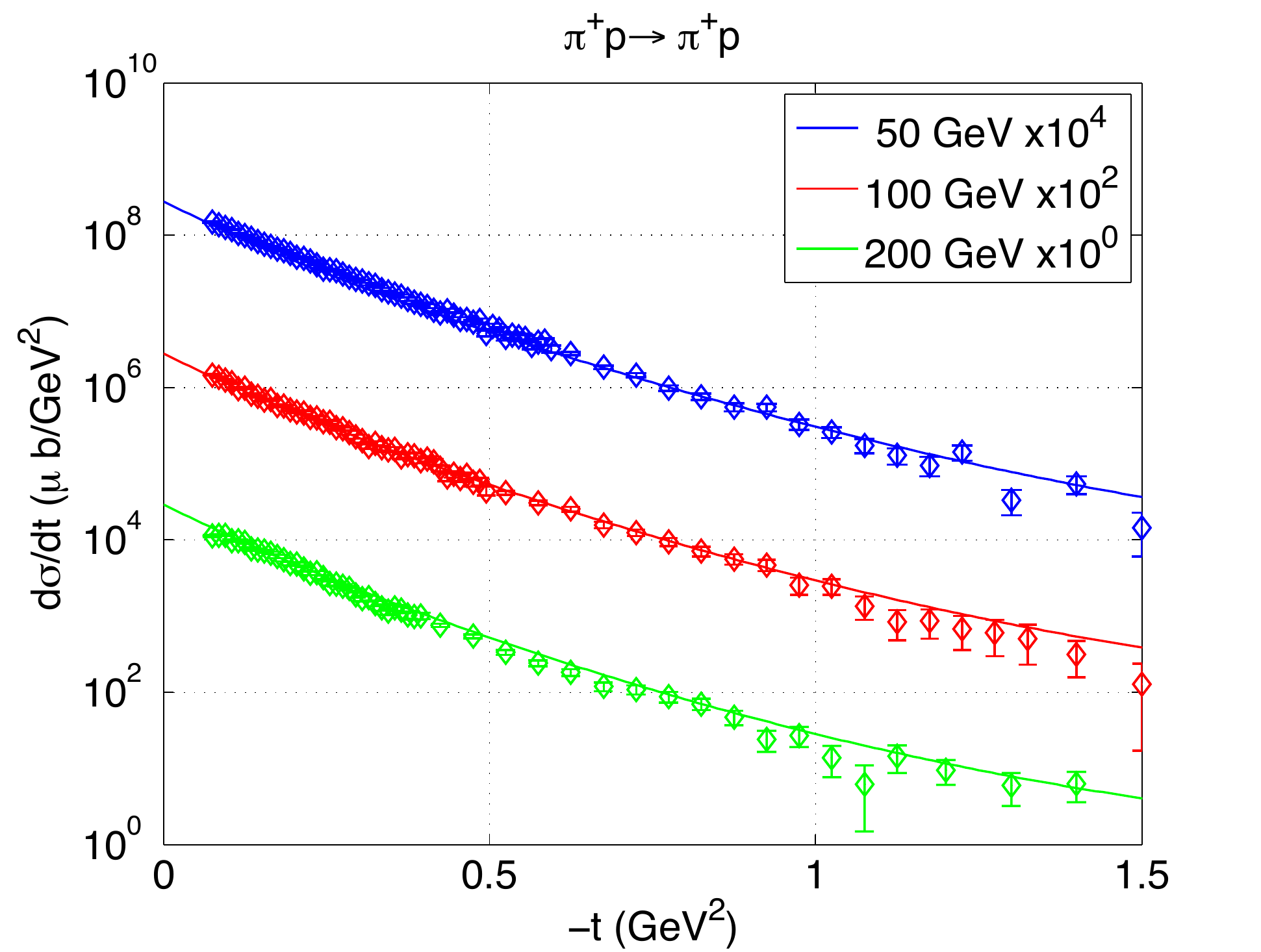}  
	 \includegraphics[width=0.3\linewidth]{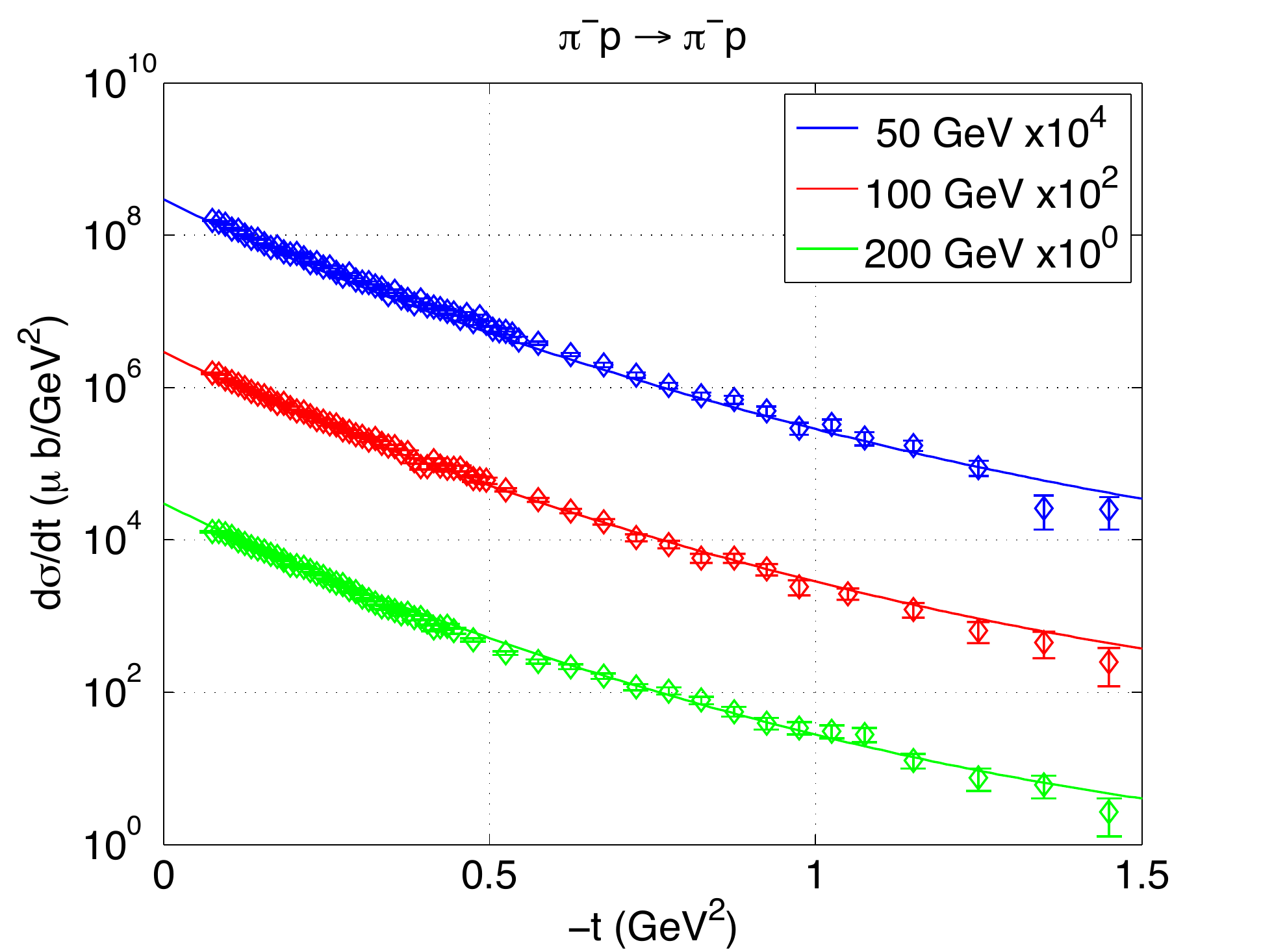} 
\caption{$\pi^+ p\to \pi^+ p$ (left) and $\pi^- p\to \pi^- p$ (right) differential cross sections for $p_{\text{lab}}=\{50,100,200\}$ GeV. Data from \cite{Akerlof:1976gk}. The contribution of the $\rho$ pole is negligible compared to isoscalar poles. \label{fig:PiP}}
\end{SCfigure}

\begin{figure}[htb]
	 \includegraphics[width=0.45\linewidth]{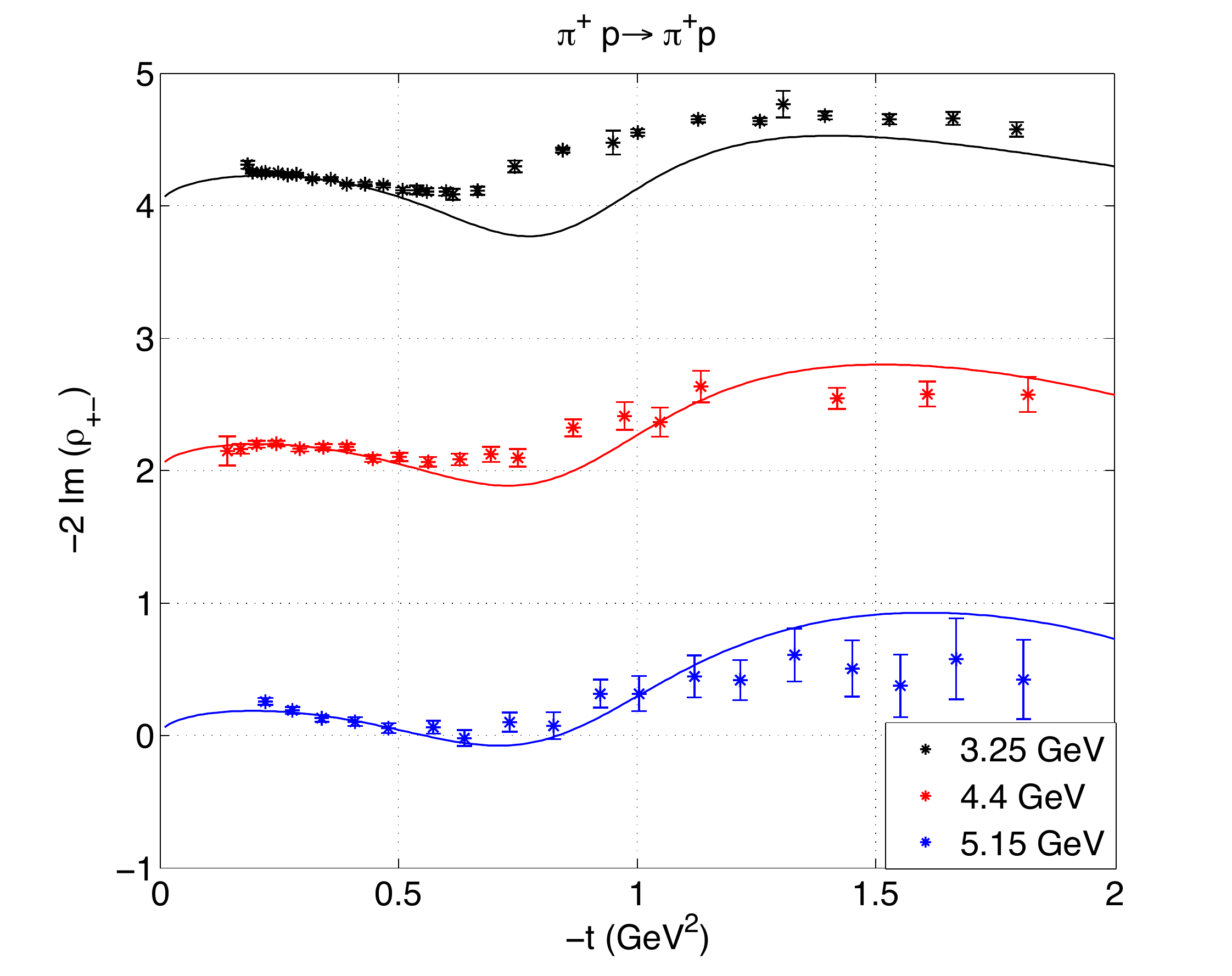}  
	 \includegraphics[width=0.5\linewidth]{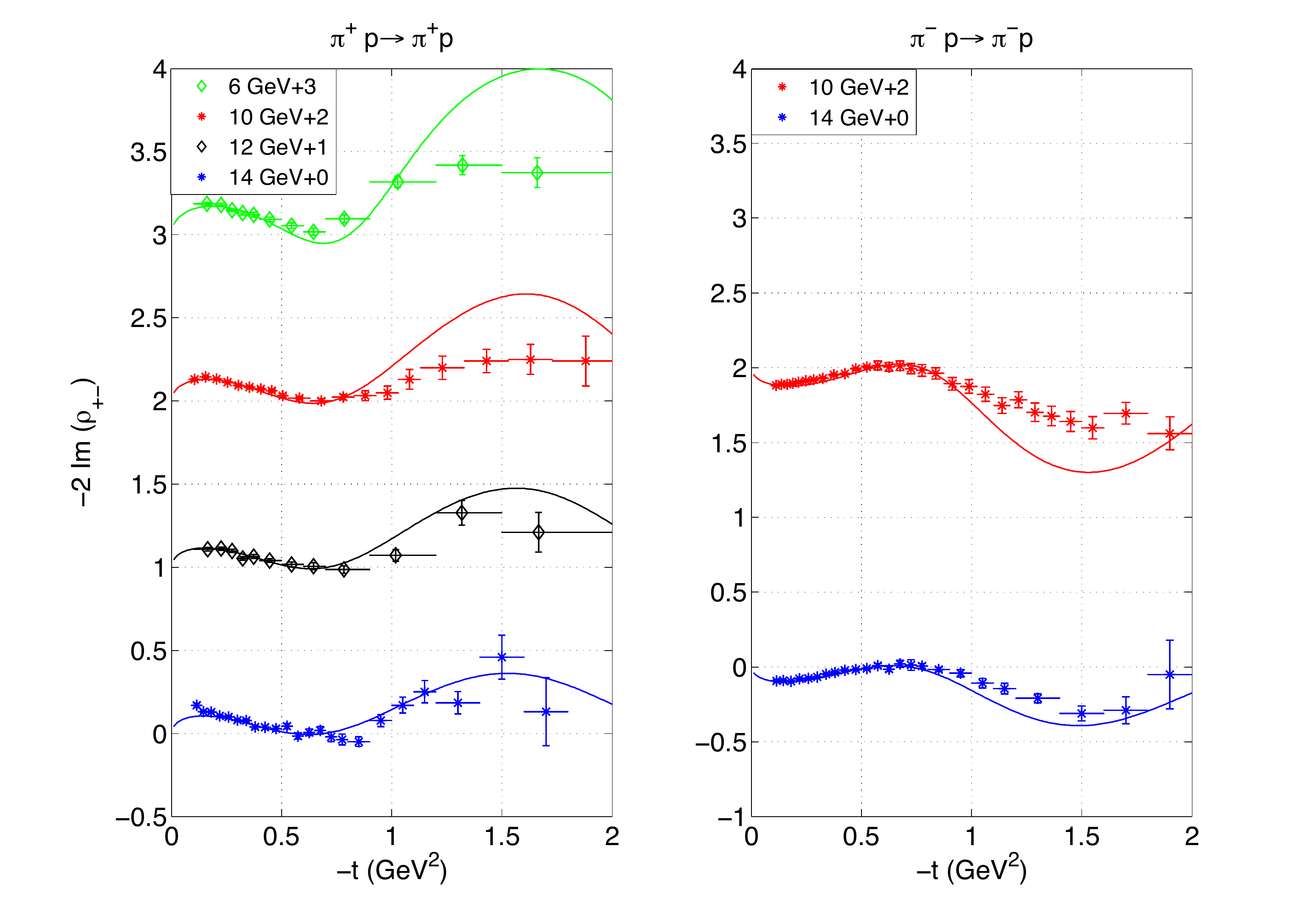} 
\caption{$\pi^\pm p\to \pi^\pm p$ analysing power for a wide range of incident momenta. Data from \cite{Fujisaki:1979fm,Borghini:1970dm,Scheid:1973si}. \label{fig:PolPiP2}}
\end{figure}

The reactions $\pi^\pm p\to\pi^\pm p$ involve ${\mathbb{P},f,\rho}$ exchanges. The parameters of the $\rho$ pole are already known. We notice that  Im($T_{++}T_{+-}^*$) are equal and opposite for $\pi^\pm p$ elastic scattering as can be seen in Fig. \ref{fig:PolPiP2}. Hence, the Pomeron and $f$ are purely non-flip at the nucleon vertex. At large energies only the Pomeron contributes to the amplitudes. The deviation from a straight lines in logarithmic plot in Fig. \ref{fig:PiP} are modeled with a small quadratic term in the Pomeron trajectory. We obtain
$
\alpha_{\mathbb{P}}(t) = 1.08 + 0.25 t + 0.15 t^2.
$
It worth mentioning that non-linearities above $50$ GeV cannot be modeled by the addition of a cut. Indeed cuts have an intercept $\sim0.5$ and are negligible at these energies in front of the Pomeron (having a bigger intercept). There is no cut associated to the Pomeron (it would be double counting). Therefore in our scheme the only possibility is to incorporate the non-linearities in the Pomeron trajectory.

\begin{SCfigure}
	 \includegraphics[width=0.65\linewidth]{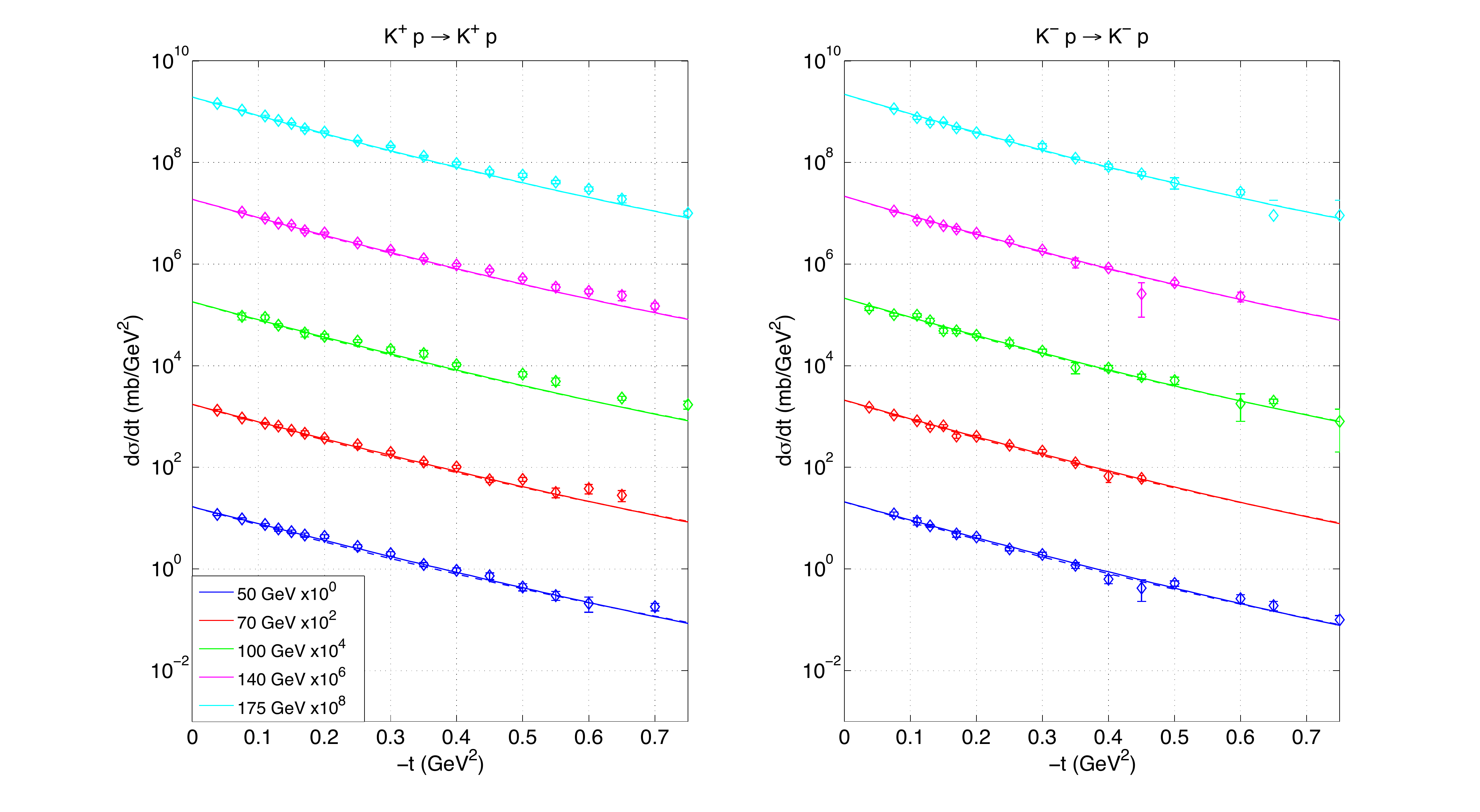}  
\caption{

$K^+ p\to K^+ p$ (left) and 

$K^- p\to K^- p$ (right) 

differential cross sections for 

$p_{\text{lab}}=\{50,70,100,140,175\}$ GeV. 

Data from \cite{Ayres:1976zm}. \label{fig:SigKP}}
\end{SCfigure}

For $K^\pm p$ elastic scattering, the situation is similar to $\pi^\pm p$ scattering. At incident momenta greater than 50 GeV the amplitude is dominated by the Pomeron. Its parameters were already determined  previously on pion-nucleon scattering and total cross sections. The model for high energy kaon-proton scattering agrees well with the data, cf. Fig. \ref{fig:SigKP} (left).  Data for $|t|>1$ GeV$^2$ at high energies would be interesting to test the non linearity of the Pomeron trajectory introduced for $\pi p$ scattering. 

\section{Nucleon-Nucleon Scattering}\label{sec:nuclnucl}
\begin{SCfigure}
	 \includegraphics[width=0.65\linewidth]{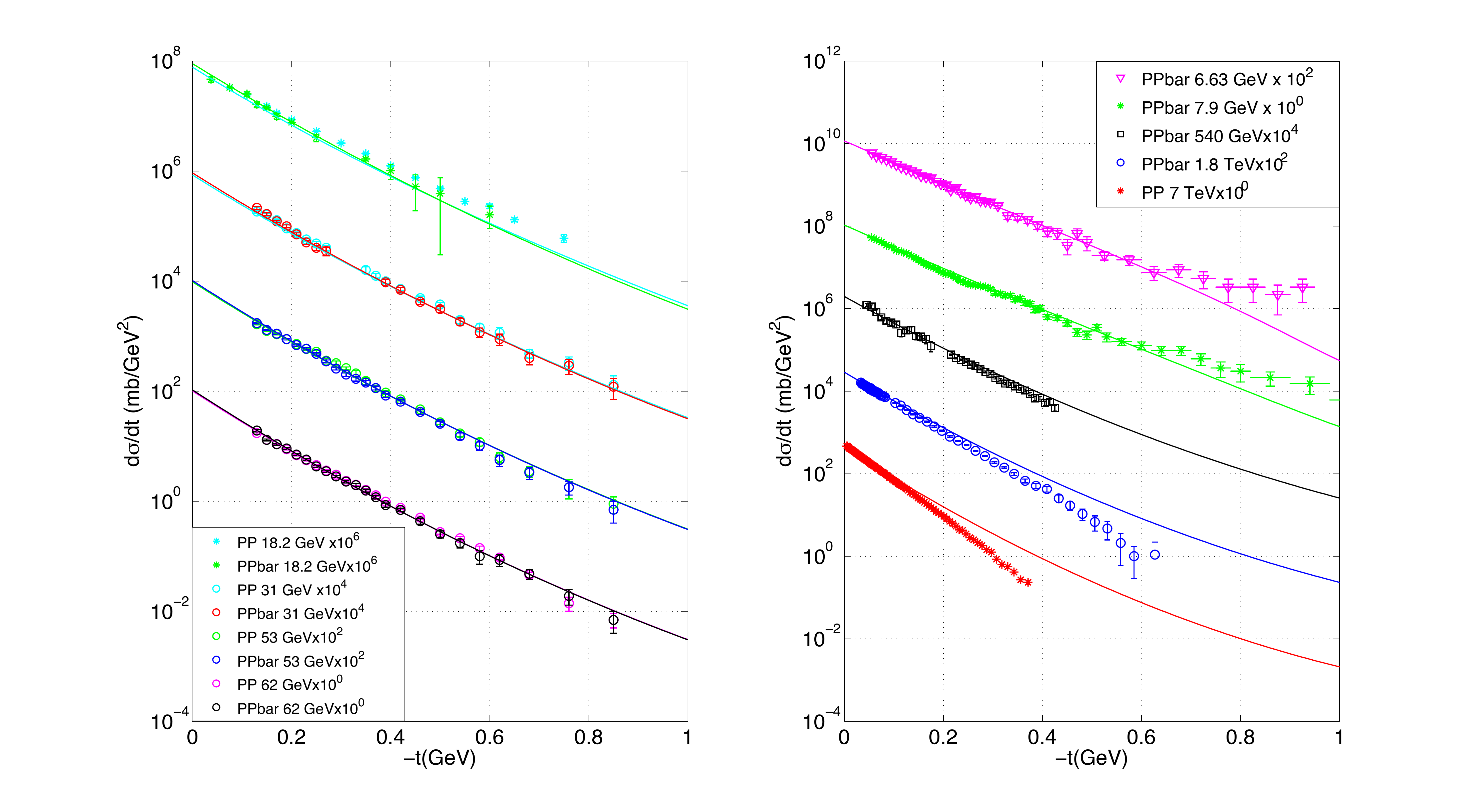}  
\caption{Comparison between $pp$ and $p\bar p$ elastic scattering (left). $pp$ and $p\bar p$ elastic scattering for a wide range of incident momenta (right). Data from \cite{Ayres:1976zm,Batyunya:1985qp,Bogolyubsky:1984ge,Amos:1990fw,Arnison:1983mm,Csorgo:2012dm}. \label{fig:SigPP}}
\end{SCfigure} 

There are 16 helicity amplitudes for the scattering of four spin $1/2$ particles. Due to parity conservation and time-reversal invariance only six of these amplitudes are independent. In addition for $NN$ scattering ($s-$channel) or $N\bar N$ scattering ($t-$channel) permutation symmetry restrict the amplitudes to only five independent. They are also only five invariant amplitudes but be consider in this work only positive naturality exchanges for nucleon-nucleon scattering (as explained they are leading trajectories). Based on a vector exchange model with $A_+$ and $B_+$ the two vector-nucleon couplings, we can decompose the amplitude according to 
\begin{align}
T &= A_+^2 (\bar v_3 \gamma_\mu u_1) ( \bar u_4 \gamma^\mu v_2) +B_+^2/(2M)^2 (\bar v_3  u_1) ( \bar u_4  v_2) 
\\
&+ A_+ B_+/(2M) \left[ (\bar v_3 u_1)  (\bar u_4 p_{13}\!\!\!\!\!\! / \ \ v_2) + (\bar v_3 p_{24}\!\!\!\!\!\! / \ \ u_1)  (\bar u_4  v_2) \right],
\end{align}
with $p_{ij}=p_i+p_j$.  We are only interested in the total and differential cross sections. The observables depend on the quantities \cite{Goldberger:1960md} (with $\bar t=t/4M^2$ a dimensionless variable)
\begin{align}
&\overline \sum_{H_i} T_{H_i}(s,t=0) = 2\nu (A_++B_+)\\
&\sum_{H_i} |T_{H_i}|^2(s,t) = (2\nu)^2 \left[ (A_++B_+)^2 - \bar t B_+^2 \right]^2
\end{align}
We associated the amplitude in the forward direction to the helicity non-flip coupling and define the dimensionless amplitudes $F^+_n=2\nu (A_++B_+)$ and $F^+_f=2\nu B_+$. This association is coherent with the effective Lagrangian for vector exchanges. Both functions $F^+_{n,f}$  involve a sum over all Regge poles and cuts [remember that for a cut there is an additional factor $\log^{-1}(\nu/\nu_0)$] and assume the form
\be
F^+_{n,f} = \sum_e [\beta^e_{+\pm}(t)]^2 \kappa(\tau_e,\alpha_e) \Gamma(j_M-\alpha_e) (\nu/\nu_0)^{\alpha_e}. 
\ee 
The magnitude of all couplings is already known. The only freedom lies in the absorption coefficient $b$ of the nucleon coupling $\beta_{NN}(0)e^{bt}$.  We find $b=2.15$ for the nucleon coupling of the Pomeron (giving $b=0.6$ for the $\pi\pi$ and $K\bar K$ couplings of the Pomeron since the sum is $2.75$ to describe properly pseudoscalar-nucleon elastic scattering).

\section{Pseudoscalar Photoproduction and Vector Hadroproduction}\label{sec:phot}
Let the masses of the pseudoscalar, vector and fermions be respectively $\mu,m, M$. The amplitudes of the reaction involving a pseudoscalar, a vector (massive or massless) and a pair of fermions can be decomposed in invariant amplitudes \cite{Berends:1967vi}, $\epsilon_\mu J^\mu= \sum_i A_i M_i $. The functions $A_i$ are free of singularities. If we impose gauge invariance, {\it i.e} $k_\mu J^\mu=0$, there are only four invariant amplitudes in the massless case $k^2=0$ and six in the general case $k^2=m^2\neq 0$.  We define four dimensionless functions free of singularities
\begin{subequations}
\begin{align}
F^+_0 & =\nu(-A_1+2MA_4)= \sum_e \beta_{10}^e(t) [\beta^e_{++}(t)-\bar t \beta^e_{+-}(t)] \kappa(\tau_e,\alpha_e)  \Gamma(j_M-\alpha_e) (\nu/\nu_0)^{\alpha_e},  \\
F^+_1&=\nu(2M A_1-t A_4)= \bar t  \sum_e  \beta_{10}^e(t) [\beta^e_{+-}(t)- \beta^e_{++}(t)] \kappa(\tau_e,\alpha_e)  \Gamma(j_M-\alpha_e) (\nu/\nu_0)^{\alpha_e},\\
F^-_0 & = \nu(A_1+t A_2)=\sum_e  \beta_{10}^e(t)  \beta^e_{++}(t) \kappa(\tau_e,\alpha_e)  \Gamma(j_M-\alpha_e) (\nu/\nu_0)^{\alpha_e},\\
F_1^- &= 2M\nu A_3 = \bar t \sum_e  \beta_{10}^e(t) \beta^e_{+-}(t) \kappa(\tau_e,\alpha_e) \Gamma(j_M-\alpha_e) (\nu/\nu_0)^{\alpha_e}.
\end{align}
\end{subequations}
Vectors contribute to positive naturality amplitudes $F^+_{0,1}$ and axial-vector contribute to $F^-_{0,1}$.  The couplings $\beta$ are introduced by matching with effective Lagrangians for vector and axial-vector exchanges. 

\begin{figure}[h]
\begin{center}
	\includegraphics[width=0.45\linewidth]{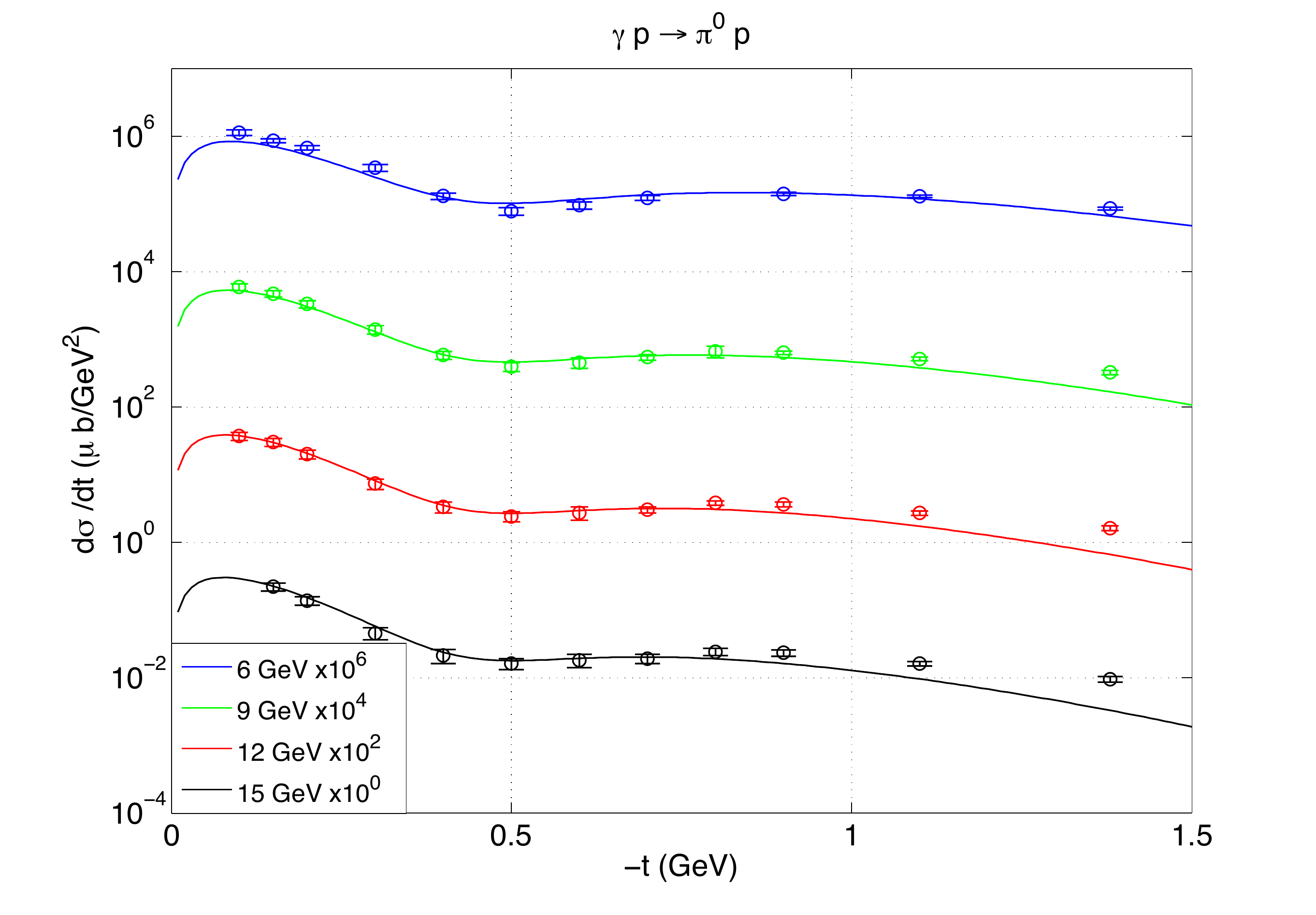}
	 \includegraphics[width=0.4\linewidth]{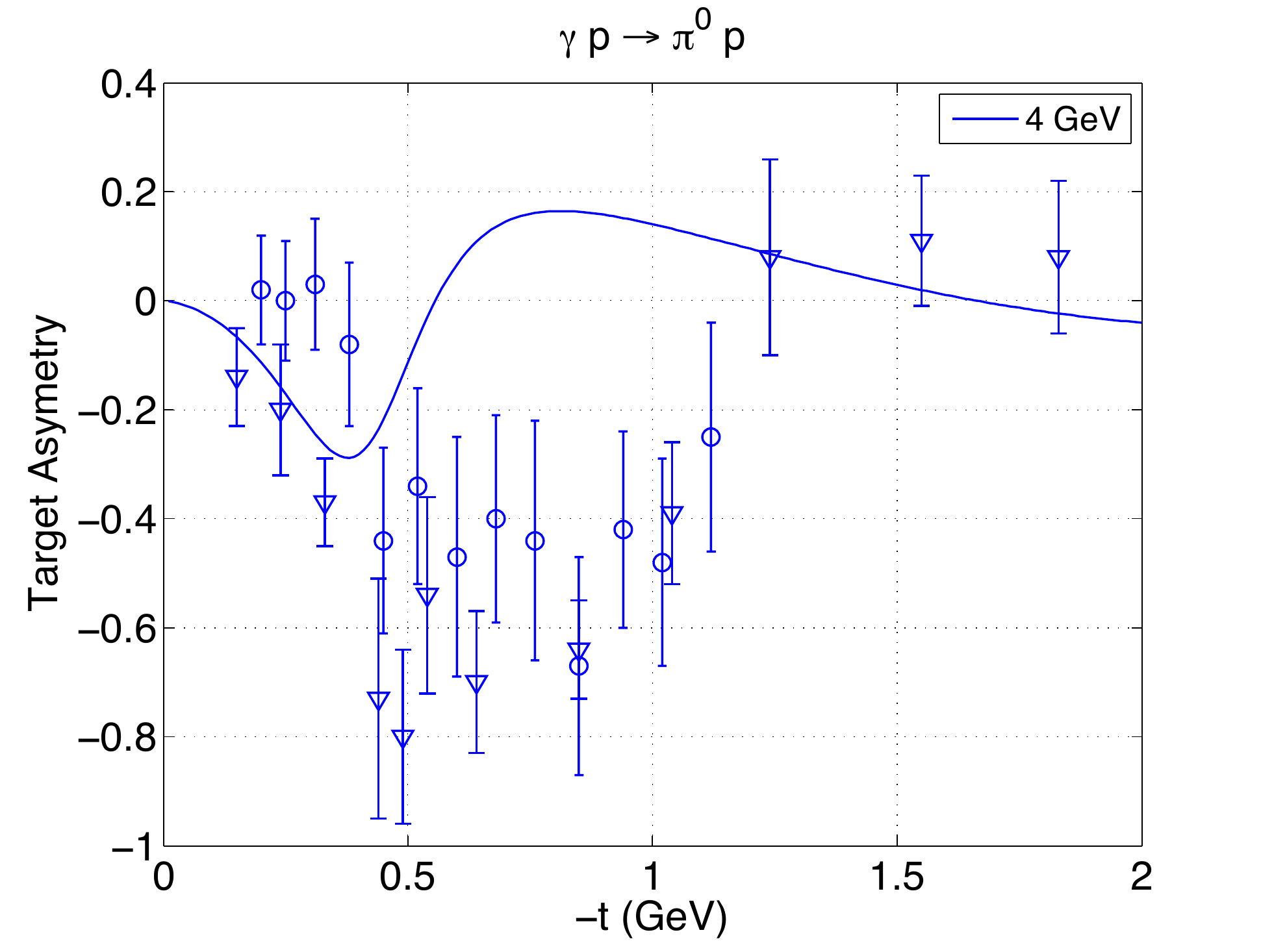}
\end{center}
\caption{$\gamma p\to \pi^0 p$ differential cross section (left) and target asymetry (right). Data from \cite{Anderson:1971xh,Booth:1972qp,Bienlein:1973pt}. \label{fig:SigPhotPi0}}
\end{figure}

\begin{SCfigure}
	 \includegraphics[width=0.7\linewidth]{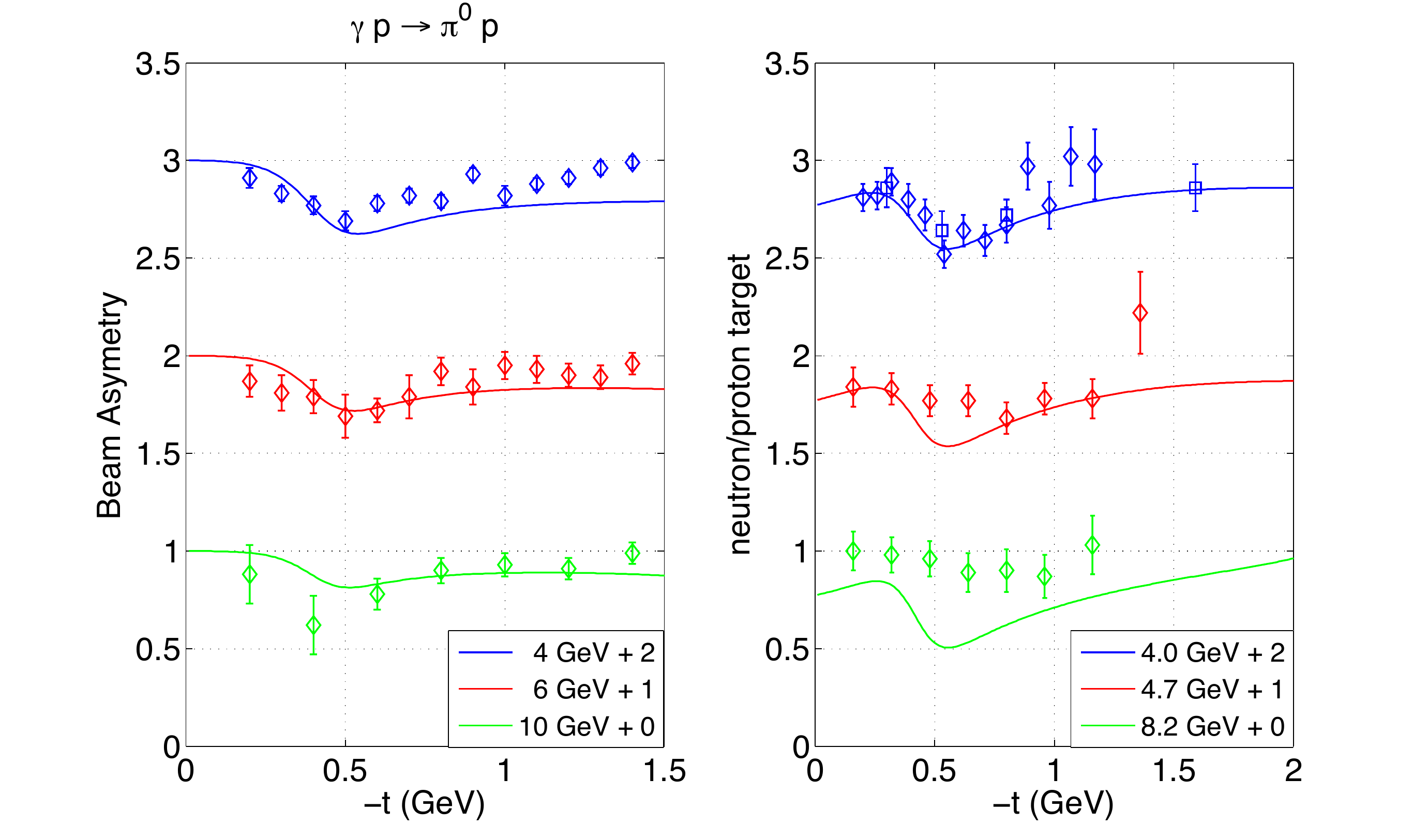}
\caption{$\gamma p\to \pi^0 p$ beam asymetry (left) and ratio $\sigma(\gamma n\to \pi^0 n)/\sigma(\gamma p\to \pi^0 p)$ (right). Data from \cite{Anderson:1971xh,Osborne:1973ed,Bolon:1971ac}. \label{fig:PolPi0}}
\end{SCfigure}

\begin{figure}[h]
\begin{center}
	\includegraphics[width=0.45\linewidth]{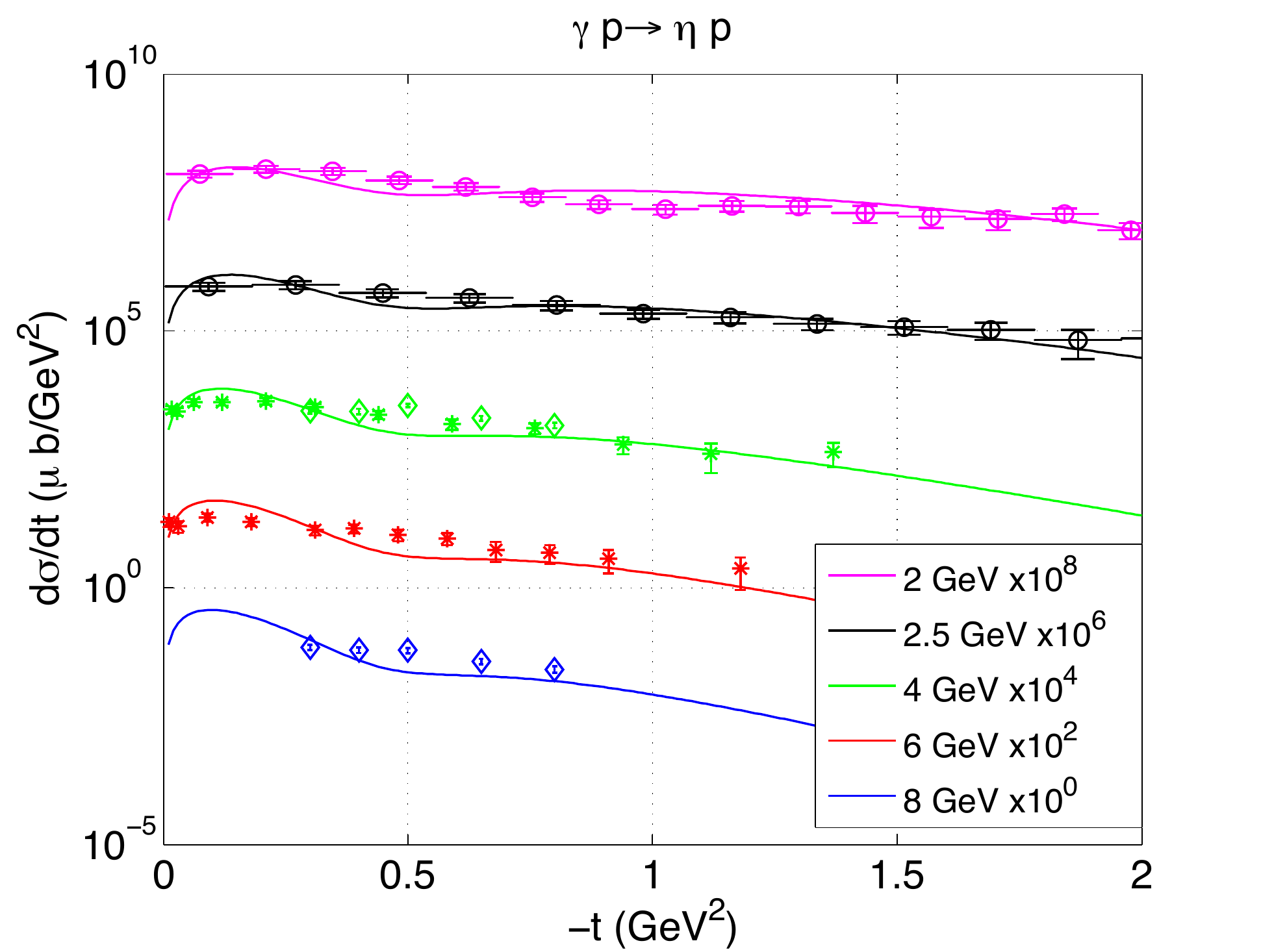}
	 \includegraphics[width=0.4\linewidth]{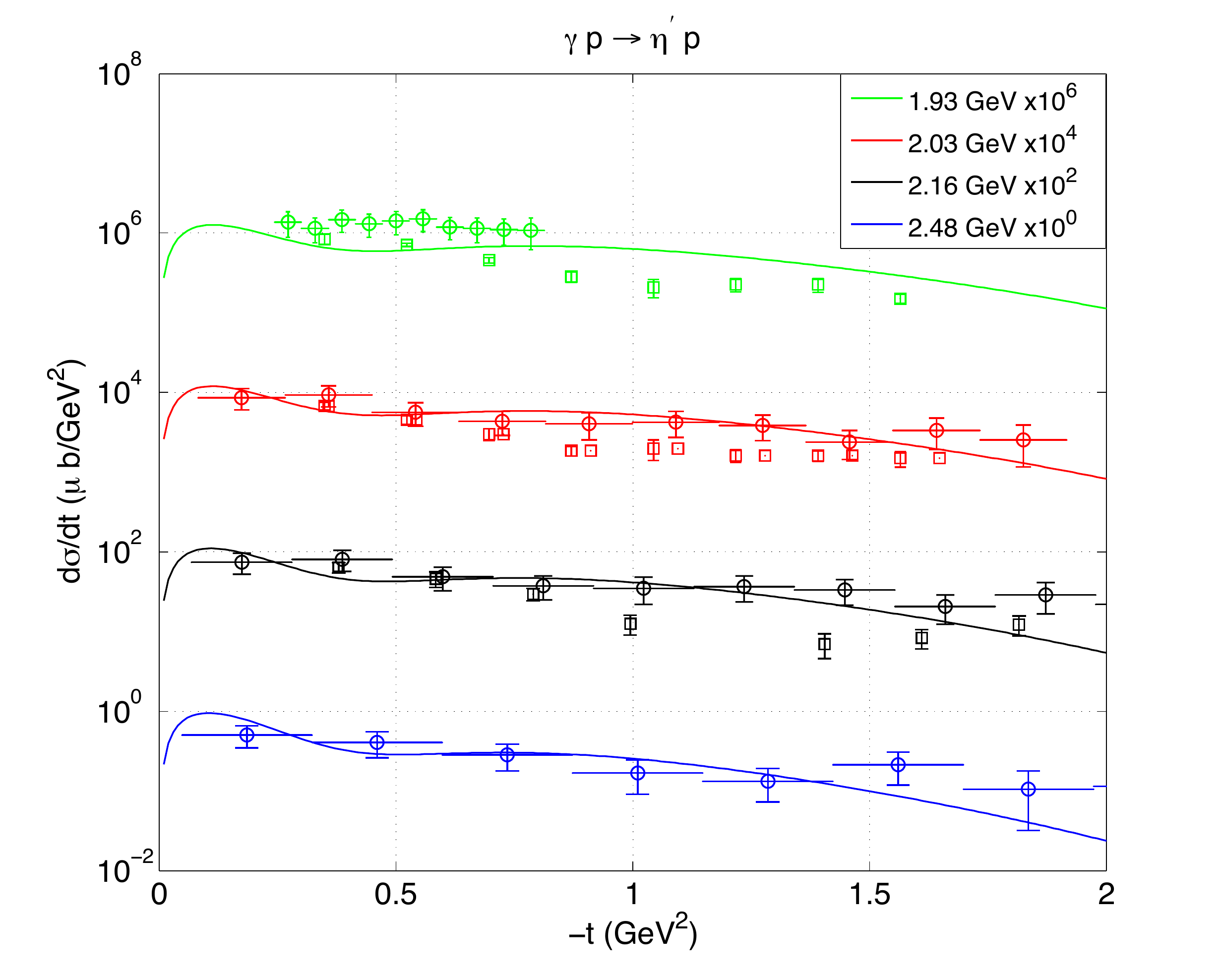}
\end{center}
\caption{$\gamma p\to \eta p$ (left) and $\gamma p\to \eta' p$ (right) differential cross section. Data from \cite{Crede:2009zzb,Dewire:1972kk,Dugger:2005my}. \label{fig:SigPhotEta}}
\end{figure}

\begin{figure}[h]
\begin{center}
	\includegraphics[width=0.4\linewidth]{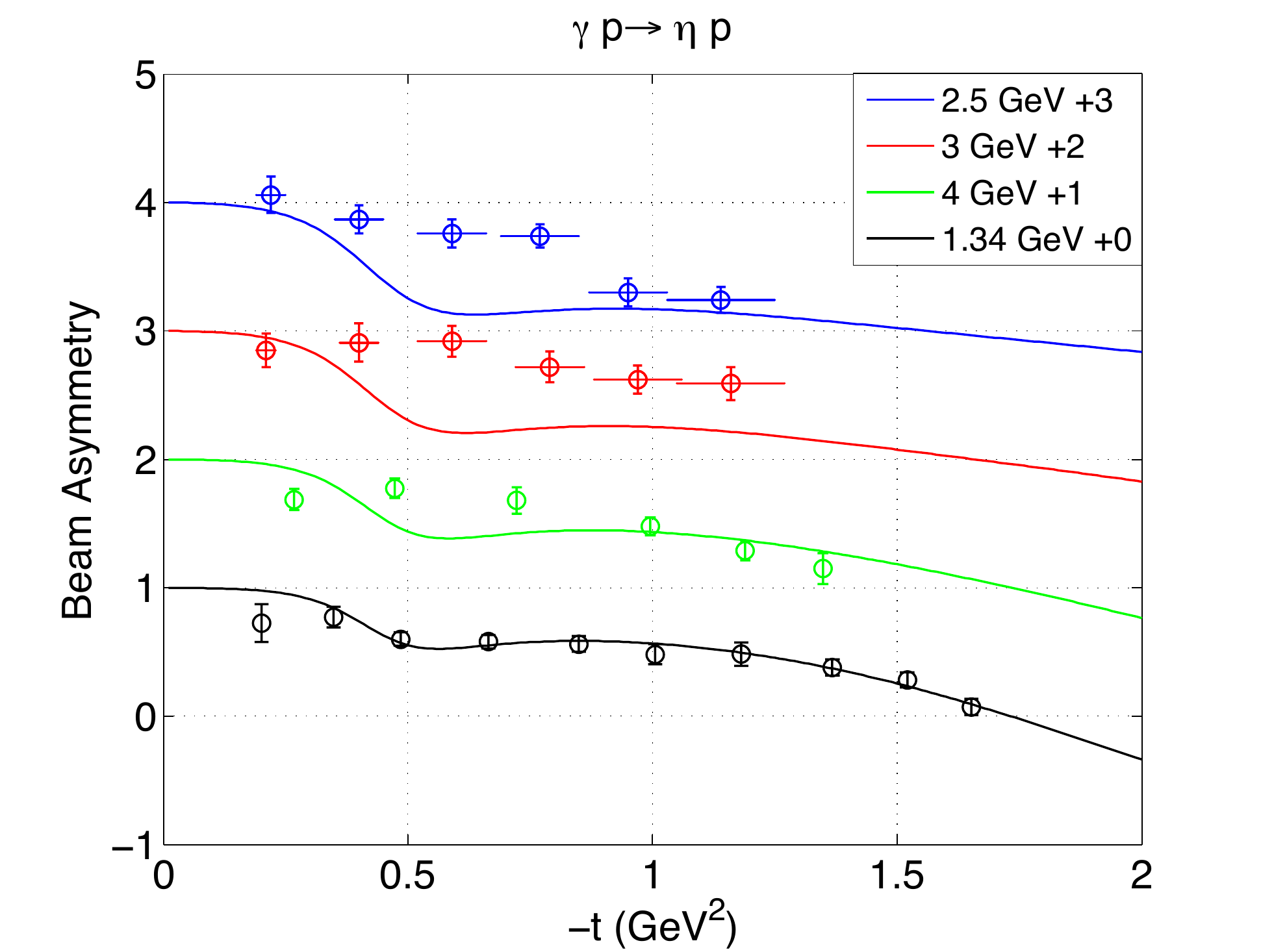}
	 \includegraphics[width=0.4\linewidth]{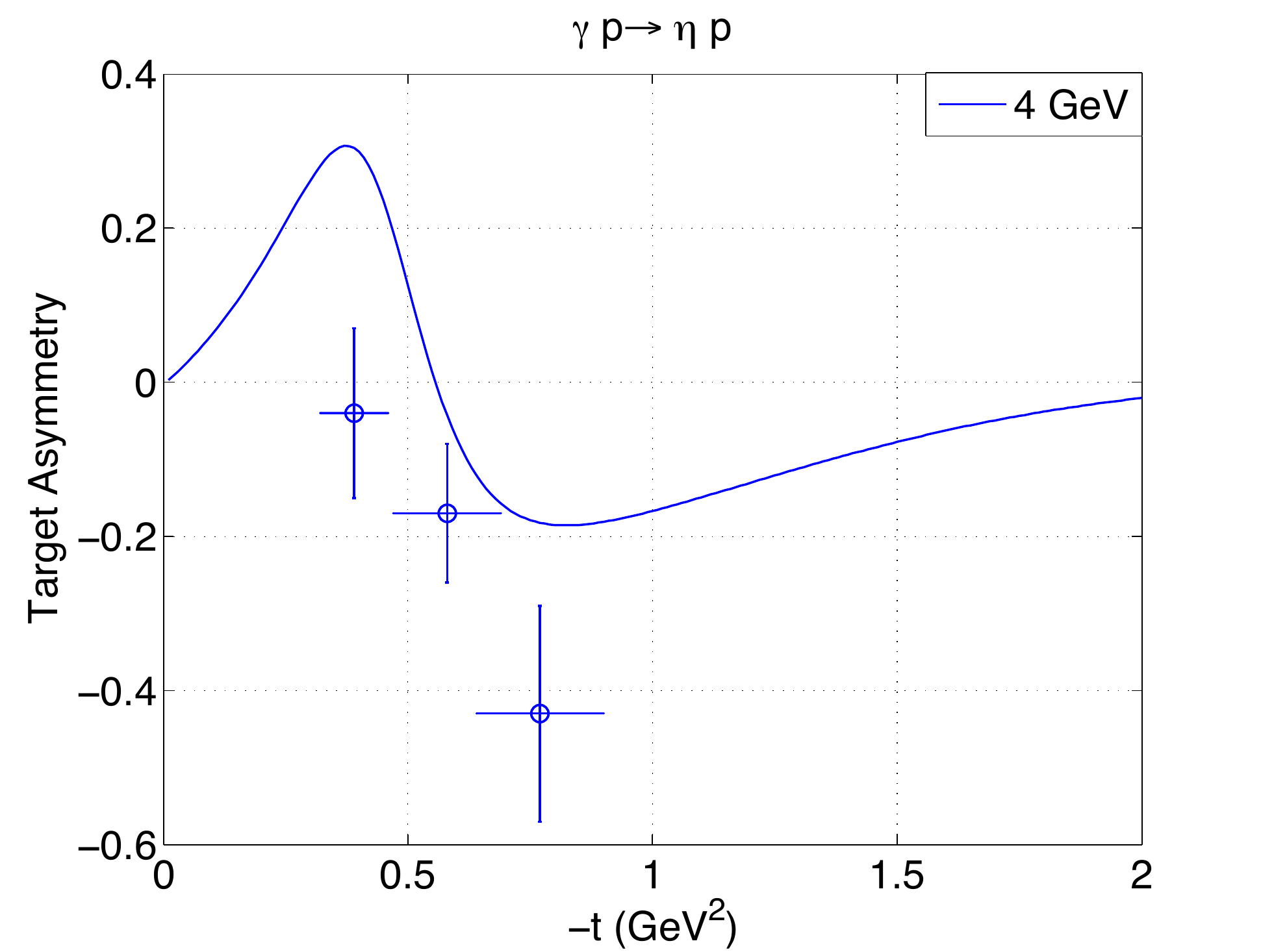}
\end{center}
\caption{$\gamma p\to \eta p$  beam (left) and target (right) asymmetries. Data from \cite{Bartalini:2007fg,Elsner:2007hm,Bussey:1976si,Bussey:1980mz}.  \label{fig:PolPhotEta}}
\end{figure}

The leading order in the energy of the observables (in $\mu$b.GeV$^{-2}$) are (remember the definition of the dimensionless variable $\bar t=t/4M^2$)
\begin{subequations}\label{eq:sig}
\begin{align}
 \frac{d\sigma}{dt} &=  \frac{389.3}{32\pi F_I^2} \left[\frac{|F_1^+|^2-\bar t|F_0^+|^2}{1-\bar t} + |F_1^-|^2 - \bar t |F_0^-|^2 \right], \\
\Sigma \frac{d\sigma}{dt} &=  \frac{389.3}{32\pi F_I^2} \left[\frac{|F_1^+|^2-\bar t|F_0^+|^2}{1-\bar t} - |F_1^-|^2 + \bar t |F_0^-|^2 \right],\\
T \frac{d\sigma}{dt} &=  \frac{389.3}{32\pi F_I^2} \sqrt{-\bar t}\ \text{Im}\,\left[\frac{F_1^+F_0^{+*}}{1-\bar t} - F_1^-F_0^{-*} \right].
\end{align}
\end{subequations}
$\Sigma=(\sigma_\bot-\sigma_\parallel)/(\sigma_\bot+\sigma_\parallel)$ is the beam asymmetry. This observable is sensitive to the negative naturality of the exchanged trajectory. The target asymmetry $T$ is the analog for photoproduction of the polarization observables Im($\rho_{+-}$). 
We compare the model with data for $\gamma p \to (\pi^0,\eta,\eta') p$ involving vector-like $(\rho,\omega)$ and axial-vector-like $(b,h)$ trajectories. The $\gamma p \to \pi^0p$ differential cross sections presents a hollow around $-t\sim 0.5$ GeV$^2$ characteristic of the vector trajectories. The same phenomenon appears in the beam asymmetry showing the importance of the axial-vector exchanges. We neglect the $h$ pole contribution and fit the parameters of the differential cross section. A good agreement with the data is achieved with the adjunction of cuts. 

The beam asymmetry displayed  in Fig. \ref{fig:PolPi0} is rater constant as the energy increases. The model is getting flatter as the energy increase simply because the intercept of the $b$ trajectory is smaller than the one of the $\rho$. On a neutron target, isovector exchanges flip sign. We assumed the same intercept for the $\rho$ and the $\omega$ trajectories. Then the ratio neutron/proton target is constant with the energy. The data show the opposite behavior indicating a possible lower intercept for the $\rho$ than for the $\omega$. 

The reaction $\gamma p\to \eta p$ is similar to neutral pion photoproduction. It involves the same exchanges and only the couplings change. Under the quark model we could predict the value of those couplings knowing the $\eta-\eta'$ mixing angle that we have determined previously. It is quite surprising that the differential cross section and the beam asymmetry are not as intuitive as in the neutral pion case. The $t-$dependence of the differential cross section is relatively flat. 
\begin{SCfigure}
	 \includegraphics[width=0.3\linewidth]{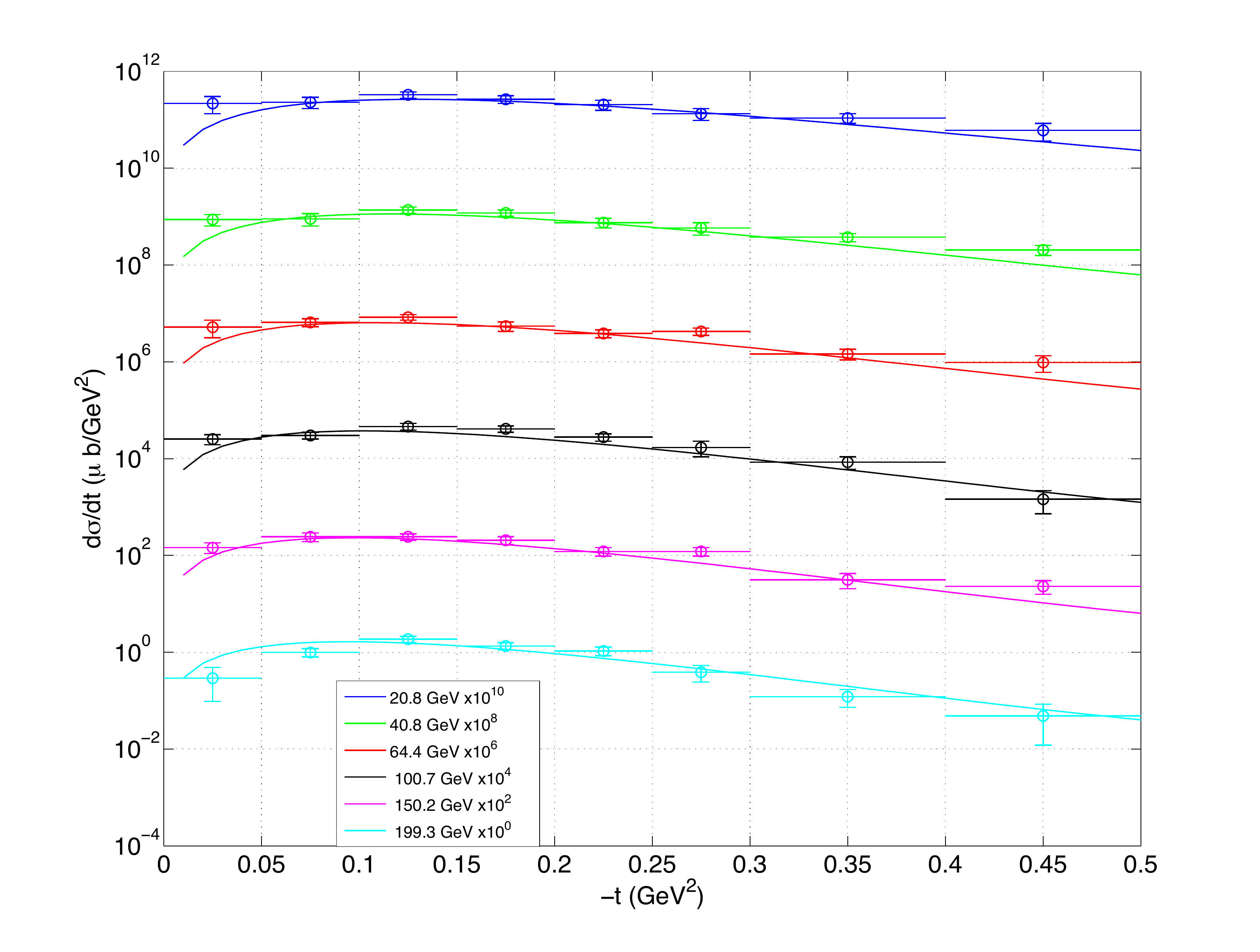}
	 \includegraphics[width=0.3\linewidth]{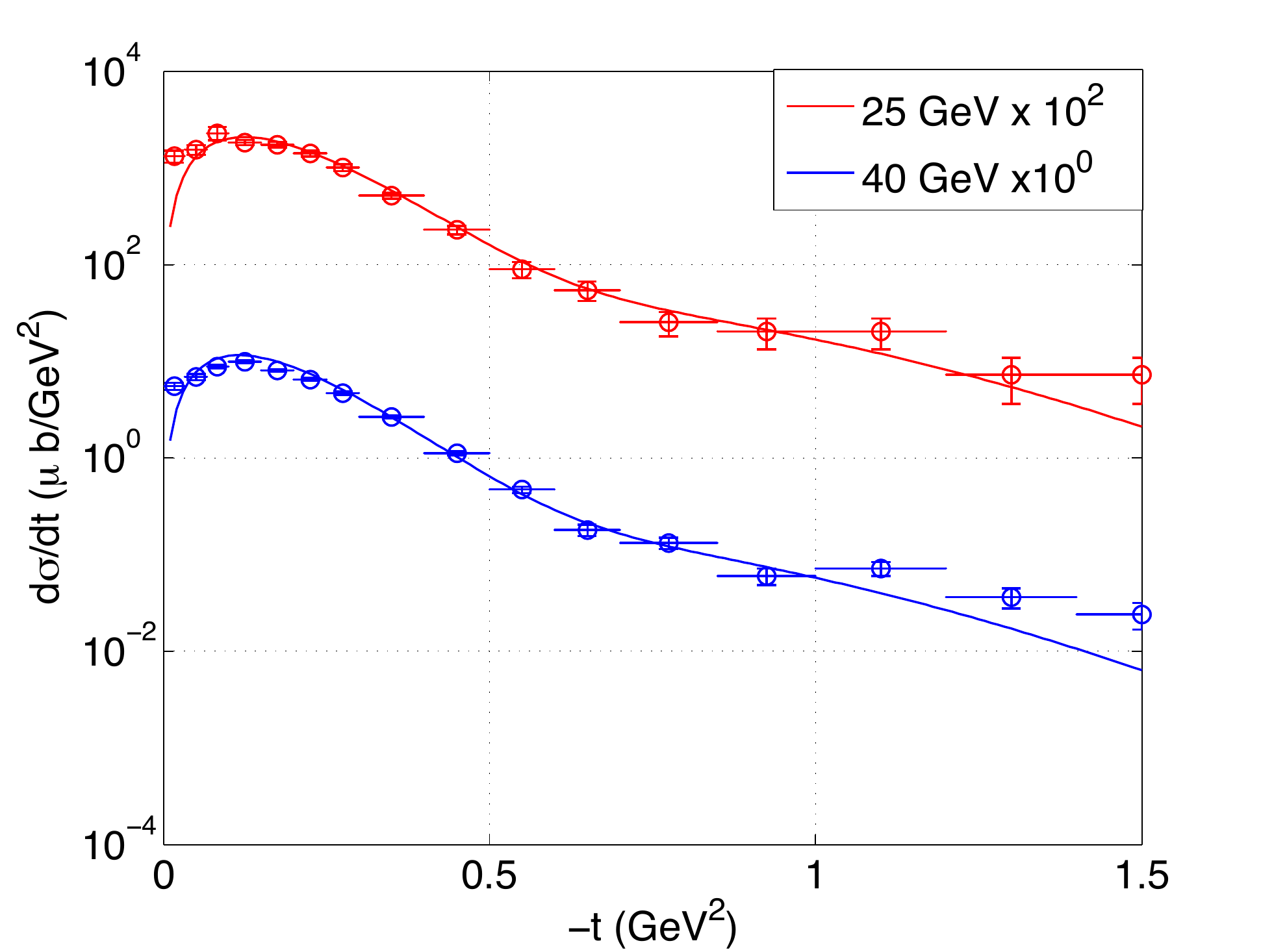} 
\caption{

$\pi^- p\to \omega n$ 

differential cross section. 

Data from \cite{Dahl:1976ky,Apel:1979yx}.  \label{fig:SigOme}}
\end{SCfigure}

\begin{SCfigure}
	 \includegraphics[width=0.65\linewidth]{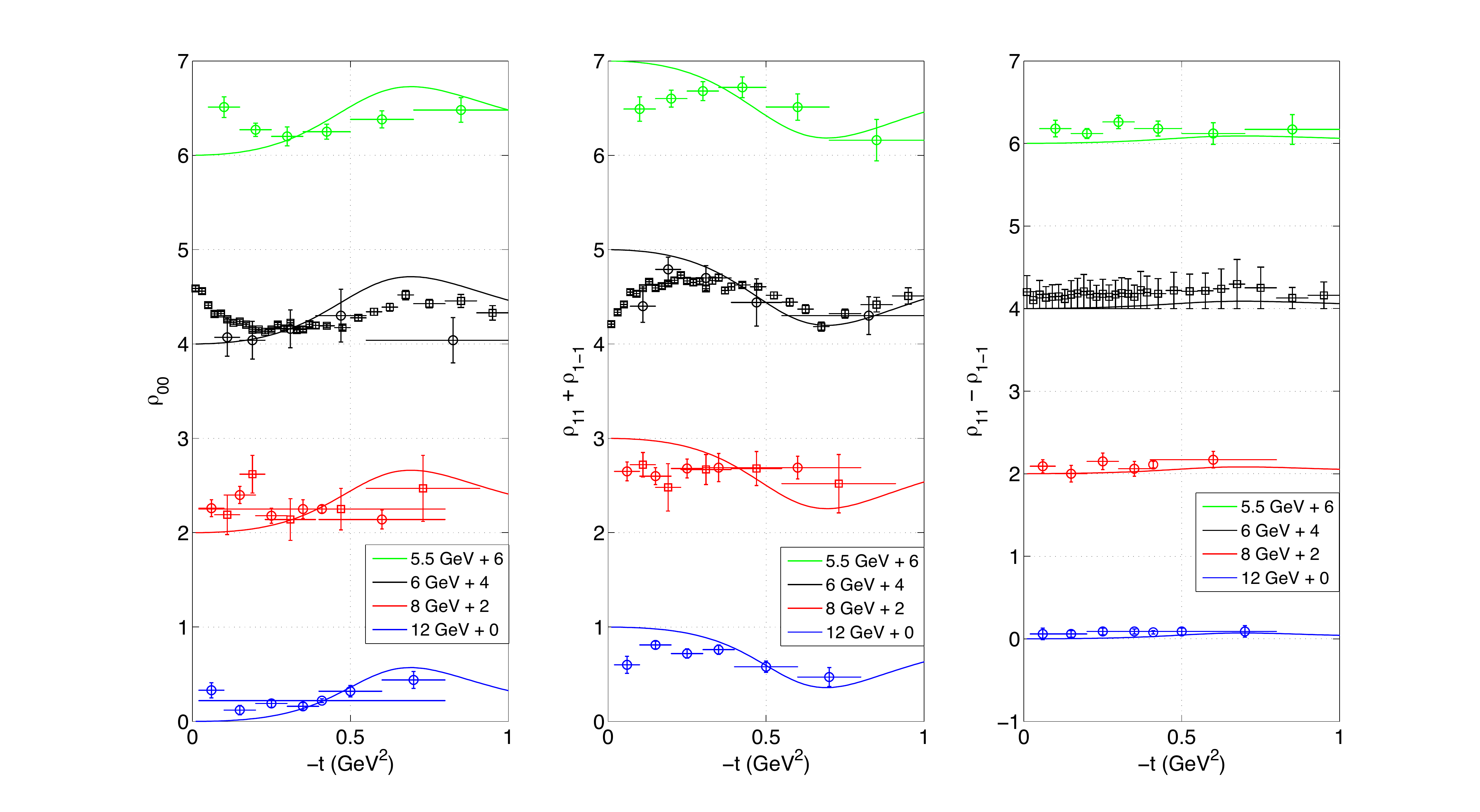}
\caption{

$\pi^- p\to \omega n$ 

spin density matrix. 

Data from \cite{Holloway:1974cp,Shaevitz:1975ir,Dowell:1975ri}. \label{fig:PolOme}}
\end{SCfigure}

For hadroproduction of a vector meson the observables are computed in term of the invariant amplitudes $F^\pm_{0,1}$ similarly to \eqref{eq:sig}. The differential cross section and the spin density matrix elements are displayed in Fig. \ref{fig:SigOme} and Fig. \ref{fig:PolOme}. We use the notation $\rho^\pm = \rho_{11}\pm \rho_{1-1}$. Of course we have $\rho^++\rho^-+\rho_{00}=1$ and only the negative naturality exchanges contribute to the longitudinal components of the vector. 

\section{Conclusion}\label{sec:concl}
We have presented a comprehensive analysis of 2-to-2 reactions involving $(f,\omega,\rho,a,b)$ exchanges. Good agreement with the data is achieved with the pole approximation but cuts are required in some cases. Residues factorize into a coupling at the beam vertex and a coupling at the target vertex. Extracting theses couplings is a first step toward a parametrization for multiple mesons production.

\end{document}